\documentclass[aps,prl,floatfix,superscriptaddress,unsortedaddress,twocolumn]{revtex4}
\usepackage[latin1]{inputenc}
\usepackage{amsmath,amssymb}
\usepackage[pdftex]{graphicx}
\usepackage{float}
\usepackage{color}
\usepackage[FIGTOPCAP,nooneline]{subfigure}
\usepackage{overpic}

\providecommand{\U}[1]{\protect\rule{.1in}{.1in}}
\newcommand{\T}{{\cal T}}
\newcommand{\PP}{{\cal P}}
\newcommand{\G}{\check{\cal G}}
\newcommand{\I}{\check{\cal I}}

\newcommand{\ua}{_{\uparrow}}
\newcommand{\da}{_{\downarrow}}

\newcommand{\slPhi}{{V}}

\renewcommand{\epsilon}{{\varepsilon}}

\begin{document}
\title{Nonlocal Thermoelectric Effects and Nonlocal Onsager Relations in a Three-Terminal Proximity-Coupled Superconductor-Ferromagnet Device}
\author{P. Machon}
\affiliation{Department of Physics, University of Konstanz, D-78457 Konstanz, Germany}
\author{M. Eschrig}
\affiliation{SEPnet and Hubbard Theory Consortium, Department of Physics, Royal Holloway, University of London, Egham, Surrey TW20 0EX, United Kingdom}
\author{W. Belzig}
\affiliation{Department of Physics, University of Konstanz, D-78457 Konstanz, Germany}
\date{Received 11 May 2012}
\begin{abstract}
We study thermal and charge transport in a three-terminal setup consisting of one superconducting and two ferromagnetic contacts. We predict that the simultaneous presence of spin filtering and of spin- dependent scattering phase shifts at each of the two interfaces will lead to very large nonlocal thermo- electric effects both in clean and in disordered systems. The symmetries of thermal and electric transport coefficients are related to fundamental thermodynamic principles by the Onsager reciprocity. Our results show that a nonlocal version of the Onsager relations for thermoelectric currents holds in a three-terminal quantum coherent ferromagnet-superconductor heterostructure including a spin-dependent crossed Andreev reflection and coherent electron transfer processes.
\end{abstract}
\maketitle

Heterostructures of ferromagnets ($F$) and superconductors ($S$) are presently subject of intense study 
since they show interesting phenomena based on the singlet-triplet conversion of pairing amplitudes at the interfaces, and the resulting spin-dependent proximity effect.
Spectacular examples are long-range triplet Josephson currents due to inhomogeneous magnetic order \cite{bergeret}, or due to the spin-dependence of the interface reflection and transmission amplitudes \cite{eschrig:2003}, that were confirmed in a set of pivotal experiments \cite{klapwijk06,sosnin06,blamire10,birge10}.
A multitude of coherence phenomena
are understood in terms of spin-dependent Andreev bound states \cite{barash:2002,huertas:2002,fogelstrom,eschrig:2003,zaikin:2007,cottet:2008a,grein:2010,eschrig:2009,metalidis:2010,zaikin:2010,linder,holmqvist}, intimately related to spin-mixing \cite{tokuyasu} and spin-filtering effects at interfaces \cite{tedrow}. 

\begin{figure}[b]
\begin{overpic}[width=1.0\linewidth]{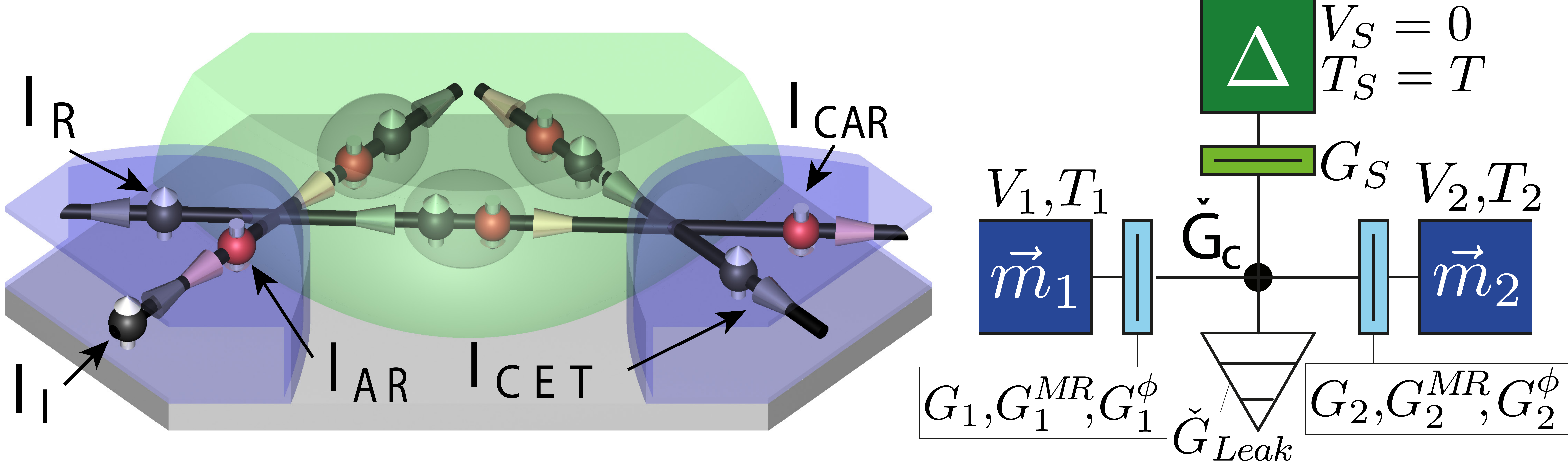}
 \put(3,27){\makebox(0,3){$(a)$}}
 \put(67,27){\makebox(0,3){$(b)$}}
\end{overpic}
\caption{\label{sys} 
(color online) $(a)$ The device consisting of two ferromagnets (to the left and right, blue) and a superconductor (in the center, green). 
Trajectories for electrons (black) and holes (red) illustrate possible transport processes in the ballistic case, as discussed in the text (white arrows denote the spin).
$(b)$ Equivalent circuit diagram of the setup shown in $(a)$ for the diffusive limit including the coherence leakage \cite{nazarov:book}. The interface parameters are discussed in detail beneath Eq.~\eqref{sbc:dirty}.
}
\end{figure}

A three-terminal superconductor-ferromagnet proximity system also 
allows to access nonlocal effects. For example, in Fig.~\ref{sys}
incoming electrons (current $I_{\rm I}$) can be reflected from the interface ($I_{\rm R}$), or enter the superconductor, where each builds a Cooper pair with another electron, leaving a hole behind that is retroreflected (so-called Andreev reflection). 
These holes can be transmitted back through the same interface ($I_{\rm AR}$), or reflected to the other interface, where they are either transmitted directly as holes ($I_{\rm CAR}$) or as electrons via the same conversion process as at the other interface in reversed order ($I_{\rm CET}$) (part of these electrons can also be reflected back to the first interface contributing to higher order processes).
Nonlocal transport has attracted considerable interest due to the latter two processes, called crossed Andreev reflection (CAR,
electron enters at one terminal and hole leaves the other terminal, or vice versa), and coherent electron transfer (CET, sometimes called `elastic cotunneling',
electron enters one terminal and electron leaves the other terminal, or the same for holes)
\cite{falci,morten,brinkman}. These processes test
the internal structure of Cooper pairs, and lead to new interesting physics that can be, and has been tested experimentally \cite{beckmann:prl,morpurgo,strunk,schoenenberger,chandrasekhar:CAR}.

In this Letter we develop 
a theory for the hitherto less explored 
nonlocal {\it thermal} transport in ferromagnet-superconductor devices, and show that a nonlocal version of Onsager relations \cite{onsager} holds in both the normal and superconducting state.  
In the superconducting state we find a strongly enhanced local thermopower and nonlocal Seebeck effect.
These effects do not require noncollinear inhomogeneities in the ferromagnetic regions or at the interfaces (a ubiquitous problem for creating triplet supercurrents \cite{bergeret,eschrig:2003,klapwijk06,braude07,blamire10,birge10}). Thus, our results should be readily observable in experiments and offer a way to  access the microscopic spin-dependent parameters.

In linear response the transport coefficients relating charge(energy) currents $I^q(I^{\epsilon})$ to an applied voltage $\Delta \slPhi_j=\slPhi_{j}-\slPhi_S$ or  temperature difference $\Delta T_j=T_{j}-T_S$
(throughout this Letter 
$j\in \left\{1,2\right\}$ labels the ferromagnet-superconductor contacts,
and $q=-|e|$ is the electronic charge) of our three-terminal system are: 
\begin{align}\label{condmat}
\begin{pmatrix}I^q_1\\I_1^{\epsilon}\\I^q_2\\I_2^{\epsilon}\end{pmatrix}=
\underbrace{\begin{pmatrix}L^{qV}_{11}&L^{qT}_{11}&L^{qV}_{12}&L^{qT}_{12}\\L^{\epsilon V}_{11}&L^{\epsilon T}_{11}&L^{\epsilon V}_{12}&L^{\epsilon T}_{12}\\L^{q V}_{21}&L^{q T}_{21}&L^{q V}_{22}&L^{qT}_{22}\\L^{\epsilon V}_{21}&L^{\epsilon T}_{21}&L^{\epsilon V}_{22}&L^{\epsilon T}_{22}\end{pmatrix}}_{\hat{L} }
\begin{pmatrix} \Delta \slPhi_1\\ \Delta T_1/T_S\\ \Delta \slPhi_2\\ \Delta T_2/T_S\end{pmatrix}\,.
\end{align}
This generalized conduction matrix $\hat L$ contains local 2x2 blocks in the diagonal, and nonlocal 2x2 blocks in the off diagonal. 
The local and nonlocal thermoelectric coefficients $L^{qT}_{ij}$ in Eq.~\eqref{condmat} give rise to large thermoelectric effects in the superconducting state, as we will show below.  
In contrast, in the normal state these coefficients are typically proportional to the asymmetry of the density of states around the chemical potential, which is orders of magnitude smaller.
Microscopically, spin-dependent scattering phases at a ferromagnetic contact 
produce an asymmetry, equal in magnitude and opposite in sign for the two spin species, in the superconducting spectrum of quasiparticles emerging from the contact. Spin filtering, which weights the spin directions differently, can resolve these asymmetric components of the spectrum. 
Both effects vanish for spin-independent systems.
Consequently, this situation is not comparable to the thermoelectric effects related to supercurrents discussed in the context of {\color{black} normal-metal/superconductor }
Andreev interferometers \cite{lambert,virtanen,titov,chandrasekhar:thermo}.
The effects we present persist also in the absence of a supercurrent emerging from the superconducting terminal.

We find that the matrix in Eq.~\eqref{condmat} (even for noncollinear magnetization configurations)
is symmetric, $\hat{L}=\hat{L}^T$,
similar to the well-known Onsager symmetries \cite{onsager}, 
however for a nonlocal setup, that contains ferromagnetic leads and includes supercurrents in the superconducting terminal as well as crossed Andreev reflection  and elastic cotunneling processes between the contacts.

We begin our theoretical analysis with the description of the interfaces between the superconductor and the ferromagnets.
Each conduction channel $n$ between a superconductor ($S$) and a ferromagnet ($F$) (with homogeneous magnetization throughout the interface region) is described by a scattering matrix
\begin{equation}
\hat {\cal S}_{n\sigma }=\left(
\begin{array}{cc}
  r_{n\sigma } e^{i\varphi^{S}_{n\sigma } } &  \quad t_{n\sigma } e^{i\varphi^{SF}_{n\sigma }}  \\
  t_{n\sigma } e^{i\varphi^{FS}_{n\sigma } }& -r_{n\sigma } e^{i\varphi^{F}_{n\sigma }}
\end{array}
\right),\,
\end{equation}
where $\sigma \in \{ \uparrow, \downarrow \}$, and
unitarity requires $r_{n\sigma}^2+t_{n\sigma}^2=1$ and 
$\varphi^{SF}_{n\sigma}+\varphi^{FS}_{n\sigma} =\varphi^{S}_{n\sigma} +\varphi^{F}_{n\sigma} $ modulo $2\pi$.
This leads
for example 
to spin-dependent conductances (spin filtering) characterized by a 
polarization $\PP_{n}=(t_{n\ua}^{2}-t_{n\da}^{2})/(t_{n\ua}^{2}+t_{n\da}^{2})$ 
and a probability for transmission,
$\T_{n}=(t_{n\uparrow}^2 + t_{n\downarrow}^2)/2$
$\le (1+|{\cal P}_n|)^{-1}$. 
Concerning the scattering phases, transport coefficients only depend on the phase shift between the reflections of spin-up and spin-down electrons on the superconducting sides of the contact, $\delta \varphi^{}_{n}=\varphi^{S}_{n\ua}-\varphi^{S}_{n\da}$, called spin-mixing angle. 
Some of the most striking consequences of the spin-dependent scattering phases are
triplet pairing \cite{eschrig:2003,eschrig:2008} or subgap resonances in the noise spectral density \cite{cottet:2008a,cottet:2008b}. 
Finally, the combination of {\it both} spin-dependent parameters $\PP_{n}$ and $\delta\varphi^{}_{n}$ leads to thermoelectric effects.
We use spin-dependent boundary conditions (SDBC) \cite{rainersauls,eschrig:2003,eschrig:2009,fogelstrom,kopueschrig,cottet} for quasiclassical Green functions in the setups shown in Fig.~\ref{sys}.

Analogously to the spin-independent theory \cite{zaitsev,nazarov:sm,nazarov:book,belzig} the system properties in the dirty limit 
(elastic mean free path much shorter than superconducting coherence length)
are fully described by the isotropic matrix Green functions $\G_{c}$ of the contact region [see Fig.~\ref{sys}$(b)$], and $\G_{j}$ ($\G_S$) for the ferromagnets (superconductor),  
that are
$8\times8$ matrices in Keldysh$\otimes$Nambu$\otimes$spin space.
$\G_{c}$ is 
determined through a finite element approach, governed by a conservation law for matrix currents \cite{nazarov:book} (see Supplementary Material for details):
$\sum\nolimits_{j}\I_{j,c}
+\I_{S,c}+ 
\I_\textrm{Leak}=0$ with the normalization condition $\G_{c}^2=1$. The leakage current $\I_\textrm{Leak}$
describes the decoherence of the superconducting order parameter due to a finite diffusion time in the central region (defining the inverse of the Thouless energy $\epsilon_{\textrm{Th}}$). The spin-dependent matrix currents ${\cal I}_{j,c}$ from contact $j$ into the superconducting contact region (denoted $c$) 
are obtained from the SDBC.
We introduce the notation $t_{n\sigma }=t_n+\sigma t'_n$ 
for spin components of the transmission quantized along a magnetization direction $\vec{m}$.
Choosing the spinor basis $\hat{\Psi}^{\dagger}=(\Psi^{\dagger}_{\uparrow},\Psi^{\dagger}_{\downarrow},\Psi^{}_{\downarrow},-\Psi^{}_{\uparrow})$
and following the line in Ref.~\cite{kopueschrig}
we find to leading order in $t_n$, $t'_n$, and $\delta \varphi_n$
a compact form for the SDBC:
\begin{equation}\label{sbc:dirty}
\I_{j,c}(\varepsilon)= \frac{q^2}{h}\sum_n
\left[  \check t_{jn} \G_{j}(\varepsilon) \check t_{jn} -i\delta \varphi^{}_{jn} \check{\kappa}_{j},\G_{c}(\varepsilon)\right] ,
\end{equation}
with 
$\check t_{jn} =t_{jn}+t'_{jn}\check{\kappa}_{j}$ and
$\check{\kappa}_{j}=\check{1}\otimes\check{\tau}_z\otimes(\vec{m}_{j}\vec{\check{\sigma}})$
($\check{\tau}$ and $\check{\sigma}$ are Pauli matrices).
The $t_{jn}$ and $t'_{jn}$ can be related to the
$\T_{jn}$ and $\PP_{jn}$ via
$(t_{jn}+ t'_{jn} \vec{m}_{j}\vec{\check\sigma})^2 ={\cal T}_{jn}\left(1+\PP_{jn}\vec{m}_{j}\vec{\check\sigma}\right)$.
Performing the sums over $n$, only few parameters remain.
In terms of the conductance quantum $G_q\equiv q^2/h$ these are 
$G_j=2G_q\sum\nolimits_n \T_{jn}$,
$G^{\rm MR}_j=G_q\sum\nolimits_n \T_{jn}\PP_{jn}$,
and $G^{\phi_{}}_j=2G_q\sum\nolimits_n\delta\varphi^{}_{jn}$,
as well as 
$\eta_{\rm Th} \equiv \varepsilon_{\rm Th} G_{S}/G_q$. Here, $G_{S}$ is the conductance between the contact region and the bulk superconductor,
fulfilling $\I_{S,c}= \frac{G_{S} }{2}[\G_{S},\G_{c}]$.
The above procedure is correct for $\delta \varphi_{jn}, {\cal T}_{jn}\ll1$, covering the full range $-1\le \PP_{jn}\le 1$. 
The equations for $\G_{c}$ are solved numerically and the density of states and the currents are calculated as function of the parameter set introduced above as described in the Supplementary Material.

In the clean limit 
(elastic mean free path much longer than superconducting coherence length)
we apply the theory developed in Refs.~\cite{eschrig:2000,eschrig:2009,zaikin:2007}. In this case, the current density 
at one particular contact can be decomposed into local (depending on the distribution function of the ferromagnet at the same contact) and nonlocal (depending on the distribution function of the ferromagnet at the other contact) contributions: incoming ($I_{\rm I}$), reflected ($I_{\rm R}$), Andreev reflected ($I_{\rm AR}$), crossed Andreev reflected ($I_{\rm CAR}$) and coherent electron transfer $I_{\rm CET}$ (see Fig.~\ref{sys}). The total current through contact $j$ into the superconductor is given by 
\begin{equation}
\label{clean0}
I^\alpha_j=I^\alpha_{j,{\rm I}}-I^\alpha_{j,{\rm R}} + I^\alpha_{j,{\rm AR}} - I^\alpha_{j,{\rm CET}} + I^\alpha_{j,{\rm CAR}}, 
\end{equation}
with $\alpha \in \{q,\epsilon\}$ and contact index $j\in \{1,2\}$.
We consider two contacts of diameter that are small compared to the superconducting coherence length $\xi_0$, and to the intercontact distance $L$. Then, quasiclassical trajectories connect the two contacts, with contact $i$ seen from contact $j$ under a solid angle $\delta \Omega_{j}={\cal A}^z_i/L^2$, where ${\cal A}^z_i$ is the area of contact $i$ projected on the plane normal to the line connecting the two contacts (here, the $z$ axis). The current through contact $j$ is proportional to ${\cal A}^z_j$, and its nonlocal part is proportional to ${\cal A}^z_1 {\cal A}^z_2/L^2$, as is the nonlocal part of the current through contact $i$.
Nonlocal contributions enter also $I_{\rm R}$ and $I_{\rm AR}$, however they are the only contributions to $I_{\rm CAR}$ and $I_{\rm CET}$. Only {\it nonlocal} contributions, via the trajectory connecting the two contacts, give rise to thermopower and Seebeck effect in the ballistic limit.

We write nonlocal current contributions as
\begin{equation}
\label{clean1}
I^\alpha_j=\frac{{\rm \delta }^2p}{{\rm \delta }\Omega}\Big|_{p_{j\rightarrow i}} \frac{{\cal A}_1^z{\cal A}_2^z}{(2\pi \hbar)^3L^2}\int_{-\infty}^{\infty} \alpha 
\left[ j_j(\varepsilon)+\tilde j_j(\varepsilon)\right] {\rm d}\epsilon ,
\end{equation}
with $({\rm \delta}^2p/{\rm \delta}\Omega)|_{p_{1\rightarrow2}}=({\rm \delta}^2p/{\rm \delta}\Omega)|_{p_{2\rightarrow1}}$ being the differential fraction of the Fermi surface 
of the superconductor with Fermi momentum such that the corresponding Fermi velocity $\vec{v}_{\rm F}$ connects the two contacts, per solid angle $\Omega $.
With the deviations of the distribution functions from that in the superconductor, for particles
$\delta f_{\rm p}$, and holes, $\delta f_{\rm h}$, the contributions to
$j_j=j_{j,{\rm I}}-j_{j,{\rm R}} + j_{j,{\rm AR}} - j_{j,{\rm CET}} + j_{j,{\rm CAR}}$
are e.g. for contact $j=1$:
$j_{1,\rm I}(\varepsilon )=2 \delta f_{1,\rm p} $,
\begin{eqnarray}
j_{1,\rm R}(\varepsilon)&=&
2|r_{1\uparrow}-v_1 t_{1\uparrow}^2r_{1\downarrow} e^{i\delta\varphi^{}_1 } \gamma_0\gamma_1 |^2 
\, \delta f_{1,\rm p}, \\
j_{1,\rm AR}(\varepsilon)&=&
(t_{1\uparrow} t_{1\downarrow})^2 
|v_1|^2 (|\gamma_1|^2+|\gamma_0|^2) 
\, \delta f_{1,\rm h}, \\
j_{1,\rm CET}(\varepsilon)&=&
(t_{1\uparrow} t_{2\uparrow})^2
|v_1u_{12}|^2 ( 1+|\gamma_0|^4 r^2_{1\downarrow} {r^2_{2\downarrow}}  )
\, \delta f_{2,\rm p},\\
j_{1,\rm CAR}(\varepsilon)&=& (t_{1\uparrow} t_{2\downarrow})^2
|v_1u_{12}|^2 |\gamma_0|^2  ({r^2_{2\uparrow}} +r^2_{1\downarrow} ) 
\, \delta f_{2,\rm h}, \qquad
\end{eqnarray}
with $\gamma_0(\varepsilon)= -\Delta /(\varepsilon+i\omega)$, $\omega(\varepsilon)=\sqrt{\Delta^2-\varepsilon^2}$, 
$\Gamma_j(\varepsilon) = \gamma_0 r_{j\uparrow} r_{j\downarrow} e^{i\delta\varphi^{}_j } $,
$u_{12}(\varepsilon)=[c-is(\varepsilon+\Gamma_2 \Delta)/\omega ]^{-1}$,
$\gamma_1(\varepsilon) = u_{12}[ \Gamma_2 c +i s(\Delta + \Gamma_2 \varepsilon)/\omega ]$,
$v_1(\varepsilon)=( 1-\gamma_1 \Gamma_1)^{-1}$,
with $c(\varepsilon)=\cosh(\omega L/\hbar {v}_{\rm F})$, $s(\varepsilon)=\sinh(\omega L/\hbar {v}_{\rm F})$. 
Finally,
$\tilde j_j(\varepsilon)$ in Eq.~\eqref{clean1} is obtained by interchanging $\uparrow \leftrightarrow \downarrow $ and $\delta\varphi^{}_j \to -\delta\varphi^{}_j$ for both contacts in the expressions above. 
The distribution functions are 
\begin{equation}
\delta f_{j,\rm p} (\epsilon ) = \frac{q\Delta \slPhi_j+\epsilon \Delta T_j/T_{S}}{4k_{\rm B}T_{S} \cosh^2(\epsilon /2k_{\rm B}T_{S})} = \delta f_{j,\rm h}(-\epsilon ).
\label{clean2}
\end{equation}
Equations \eqref{clean0}-\eqref{clean2} are valid for arbitrary transparencies and spin polarizations.
Nonlocal effects decay when $L$ exceeds the scale of the superconducting coherence length ($\xi_0=\hbar {v}_{\rm F}/k_{\rm B}T_{\rm c}$ in the clean limit). See Supplementary Material for examples.

The temperature dependence of the superconducting pair potential  $\Delta$ is taken into account by solving self-consistently the gap equation in weak coupling BCS theory (with its zero temperature value denoted $\Delta_0$). 

As shown in the Supplementary Material,
in ballistic systems
only processes that involve the opposite contact contribute to the local thermoelectric coefficients $L_{jj}^{qT}$ and $L_{jj}^{\varepsilon V}$.
The term $I_{j,{\rm AR}}^\alpha $ does not contribute because
$j_{1,{\rm AR}}(-\varepsilon) $ cancels the corresponding term for $\tilde j_{1,{\rm AR}}(\varepsilon)$ in the expressions for the thermoelectric coefficients [both have the same pre-factor $(t_{1\uparrow}t_{1\downarrow})^2$, i.e. spin filtering is not active here]. 
In contrast, the expression for 
$I_{j,{\rm R}}^\alpha $
does not show such a cancellation when contact 1 is spin polarized, due to the asymmetric combination of transmission and reflection coefficients 
in $j_{1,{\rm R}}(\varepsilon)$ 
(i.e. spin filtering is active) and the presence of spin mixing ($\delta \varphi_1$). It does, however, require in addition that $r_{2\uparrow}r_{2\downarrow} e^{i\delta \varphi_2}\ne 1$ (which means the presence of a second contact) in order for it to cause nonzero thermoelectric effects.
When the impurity mean free path or the 
dimension of the superconducting terminal shrinks below $\xi_0$, direct backscattering due to impurities or surfaces contributes and leads to a local thermopower even in a two-terminal device.

\begin{figure}[h]
\begin{overpic}[width=0.49\linewidth]{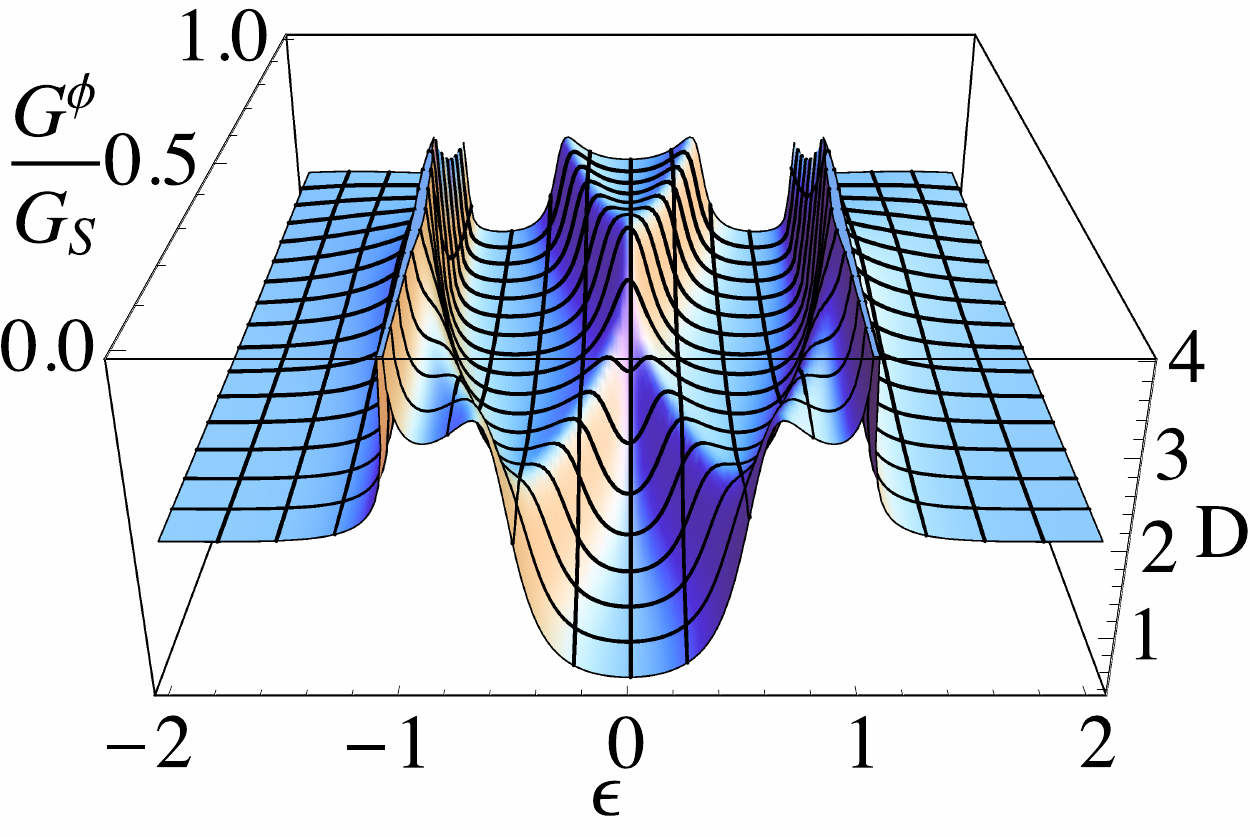}
 \put(3,66){\makebox(0,3){$(a)$}}
\end{overpic}
\begin{overpic}[width=0.49\linewidth]{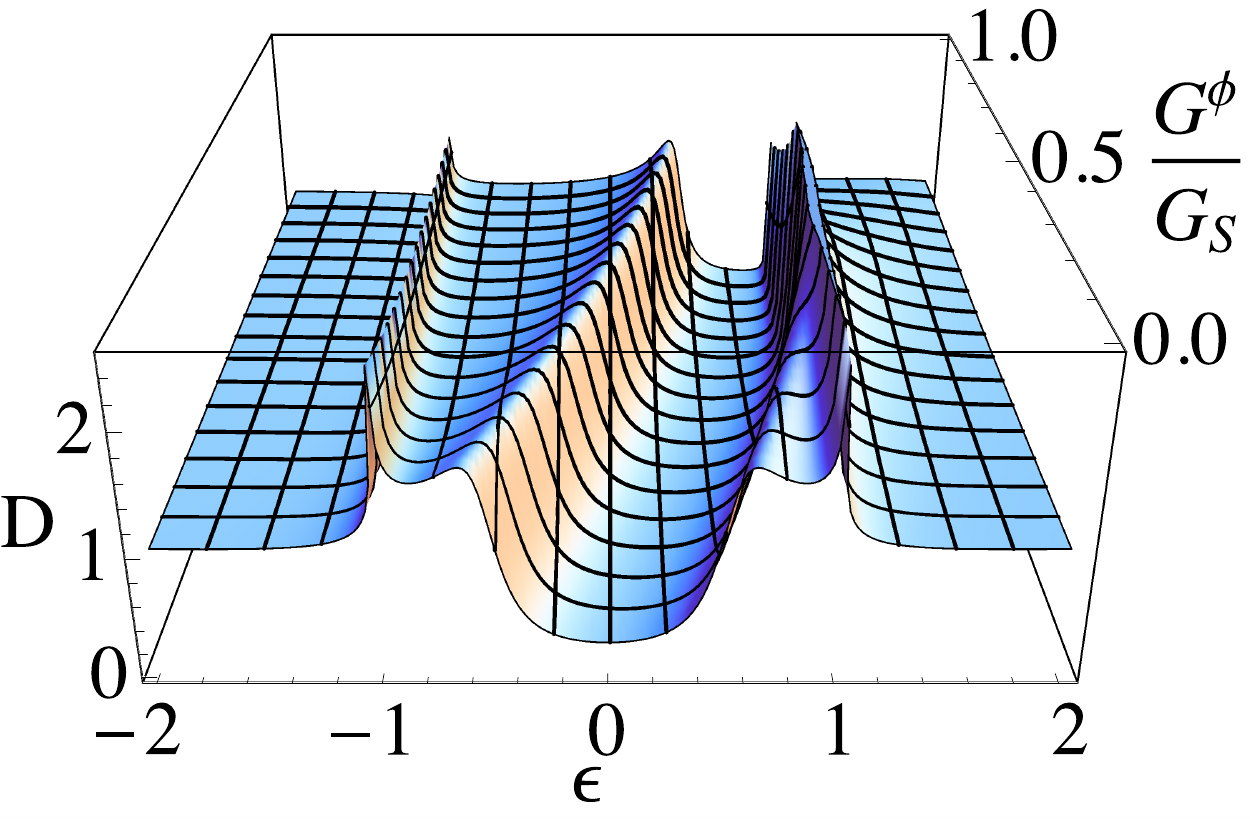}
 \put(3,66){\makebox(0,3){$(b)$}}
\end{overpic}
\caption{\label{dos} (color online) Density of states $D$ in the contact region for 
$G_1=G_2=0.1G_{S}$, $G^{\rm MR}_1=G^{\rm MR}_2=0.005 G_{S}$
(10\% polarization)
and $\eta_\textrm{Th}\equiv \varepsilon_{\rm Th} G_{S}/G_q=0.5\Delta_0$ (with the Thouless energy $\varepsilon_{\rm Th}$ of the contact region).
$(a)$ Total DOS depending on the spin-mixing term $G^{\phi}$ for equal ferromagnets. The $G^{\phi}$ term splits the pseudo gap into the different spin directions. $(b)$ shows the asymmetry in the SDOS for spin-down (the spin-up SDOS looks equal but mirrored at the $\epsilon=0$ axis).}
\end{figure}

As the mechanism behind the thermoelectric effects can be understood from 
the density of states (DOS) in the contact region, we discuss first this quantity. In the dirty limit (see Fig.~\ref{dos}) for $G^{\phi}=0$ the DOS displays peaks at $\epsilon=\Delta$ resulting from the superconducting leads and the proximity induced minigap. The magnetization directions are chosen parallel. Increasing $G^{\phi}$ simultaneously in both terminals leads to a Zeeman splitting of the minigap in spin-up and down parts and consequently breaks the symmetry of the spin-projected DOS (SDOS) around the Fermi energy $\epsilon_{\rm F}$ (see~Fig.~\ref{dos}). Hence, we expect a nonvanishing thermopower if a spin-filtering term $G^{\rm MR}$ is present simultaneously.
An equivalent discussion of the SDOS depending on the spin-mixing angle $\delta \varphi$ for a ballistic system is done in \cite{metalidis:2010}. The subgap peaks there are much sharper compared to the washed-out peak in the dirty limit.
This is associated with the fact that only trajectories connecting the two contacts contribute to the nonlocal transport, in which case it is governed by a 
single length $L$.
This is not the case in diffusive structures, where quasiparticles take random paths of various length between the contacts {\color{black} (and back to the same contact). }
Nevertheless, both ways lead to an asymmetry in the SDOS and consequently to the astonishing prediction of giant thermoelectric effects for spin-polarized interfaces.

We now turn to the  experimentally relevant question how to define a nonlocal thermopower $\mathcal{S}_{12}=-\Delta \slPhi_{1}/\Delta T_{2}$, which is not unique in contrast to the local thermopower $\mathcal{S}_j=-\Delta V_j/\Delta T_j=L^{qT}_{jj}/(T_SL^{qV}_{jj})$.
In the Supplementary Material we discuss several possibilities to relate voltage and temperature differences between the two ferromagnets and the superconductor avoiding a control of energy currents. In this Letter we chose to define the thermopower at contact 1 via ${\cal S}_{12} = L_{12}^{qT}/(T_SL_{11}^{qV})$,
which is caused by a temperature difference $\Delta T_2$ at contact 2 under the conditions $\Delta V_2=0$, $\Delta T_1=0$, $I^q_1=0$.

\begin{figure}[b]
\begin{overpic}[width=0.49\linewidth]{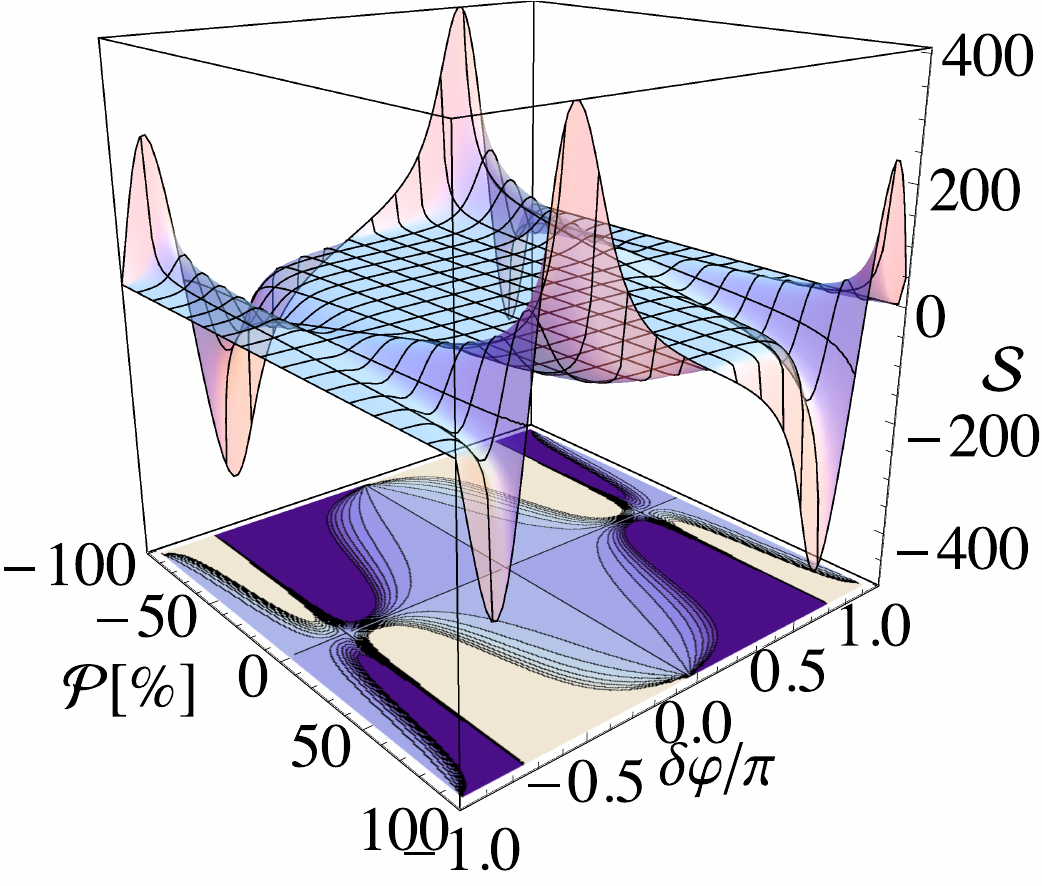}
 \put(3,85){\makebox(0,3){$(a)$}}
\end{overpic}
\begin{overpic}[width=0.49\linewidth]{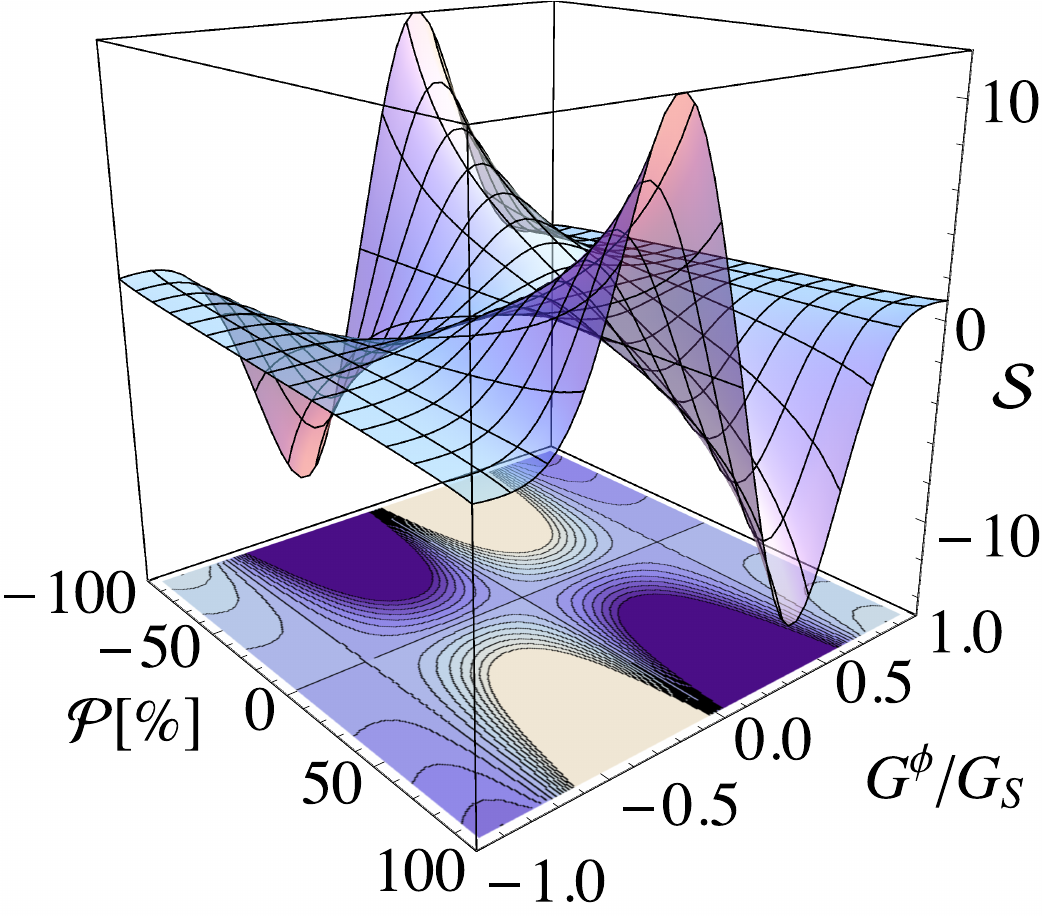}
 \put(3,85){\makebox(0,3){$(b)$}}
\end{overpic}
\caption{\label{phasendiagramm} (color online) Nonlocal thermopower 
$\mathcal{S}=L^{qT}_{12}/T_SL^{qV}_{11}$ 
for a symmetric setup as function of polarization $\PP $ and spin-mixing parameter in the clean $(a)$ and the dirty $(b)$ limit
for $T=T_S=0.1\,T_{\rm c}$.
We assume equally polarized channels, $\PP_n\equiv \PP$.  In $(a)$
$\T_{n1}\equiv \T_1 =0.1=\T_2\equiv \T_{n2}$, $L=0.5\xi_0$, $\delta \Omega_1 = \delta \Omega_2 = \pi/20$; in $(b)$ $G_1=G_2=0.1G_{S}$ and $\eta_\textrm{Th}=0.5\Delta_0$.
$\mathcal{S}$ is plotted in units of $g\, k_{\rm B}/|q|$ where
$g=-\T_2 (1+\PP^2 )\delta \Omega_2/2\pi$ in the clean limit, and $g=-G_2/(G_2+G_{S})$ in the dirty limit.
}
\end{figure}
In Fig.~\ref{phasendiagramm} we
show the dependence of $\mathcal{S}\equiv \mathcal{S}_{12}$ on the polarization and spin mixing for $T/T_{\rm c}\ll1$ assuming equal ferromagnets. 
The clean and the diffusive limit show
similar behavior, in particular for weak polarizations. 
For large polarization, values of more than 100$\mu$V/K are achievable in both limits.
Both limits exhibit
the same point symmetry with respect to the origin, and vanish if one of the spin-dependent parameters vanishes. 
This behavior is understood from the SDOS as follows. 
The symmetry of $\mathcal{S}$ with respect to the origin is according to Eq.~\eqref{sbc:dirty} a consequence of a $\pi$ rotation in spin space. The trace in the current formula (shown in the 
Supplementary Material)
is invariant under such a unitary transformation.
The sign change  with respect to the axes can be understood by Fig.~\ref{dos}. 
The two spin projections produce thermoelectric effects with opposite signs. 
Depending on positive or negative $G^{\rm MR}$ one or the other of the two contributions will be weighted more. Thus, a sign change in $G^{\rm MR}$ changes the sign of the thermopower. 
On the other hand, a sign change in $G^{\phi}$ interchanges the roles of spin-up and spin-down contributions to the DOS, and hence changes the sign of the thermopower too. 
Similar arguments explain the zero crossing of the thermopower 
when both spin-polarized peak positions
in Fig.~\ref{dos}$(a)$ cross the Fermi level. The same mechanism leads to a sign change in the clean limit, when the spin-split Andreev levels cross at the Fermi energy. Here the effect is even more drastic since the width of the crossing peaks is determined solely by the transmission to the ferromagnets. 

\begin{figure}[t]
\includegraphics*[width=0.9\linewidth,clip]{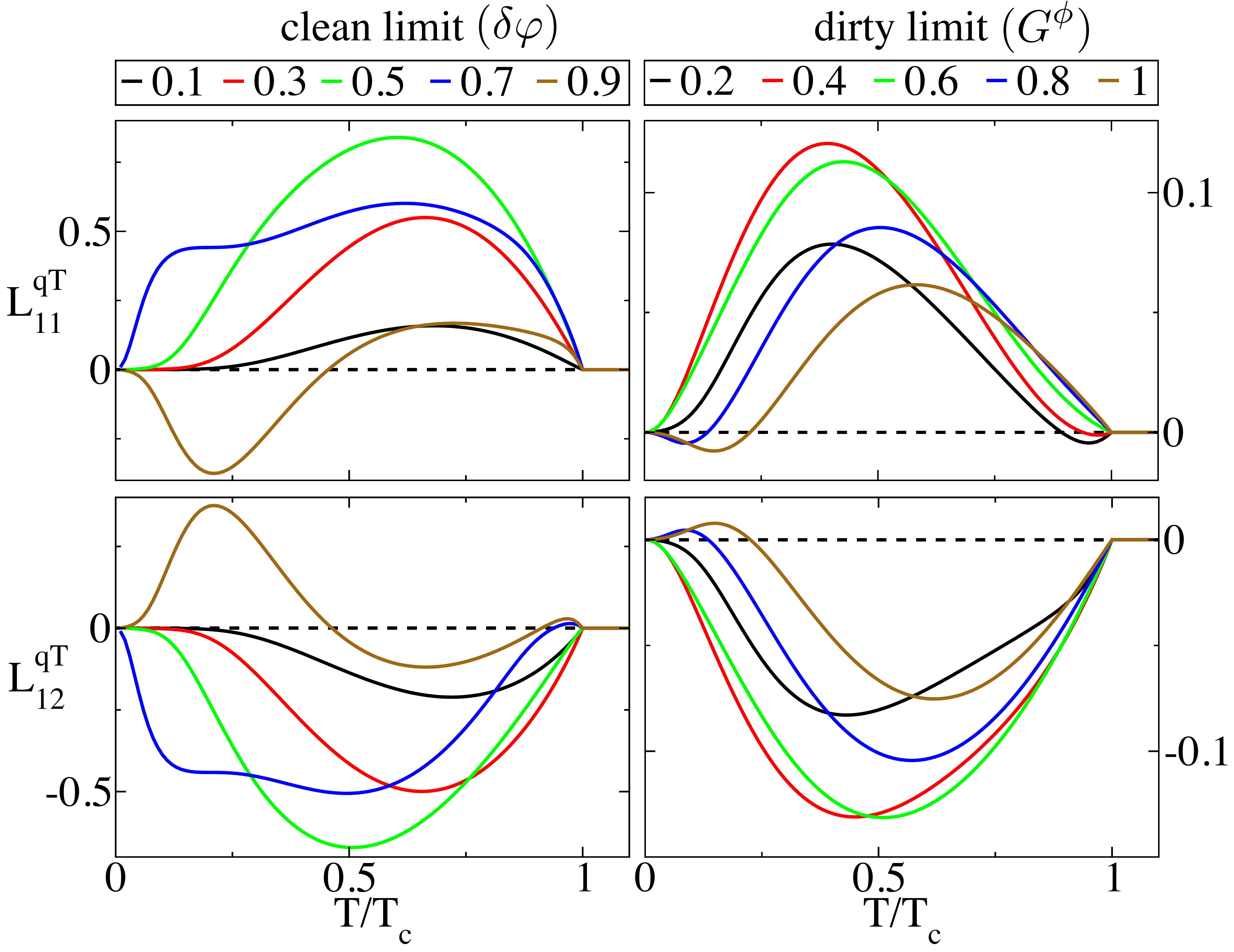}
\caption{\label{tempdep}(color online) Temperature dependence of local and nonlocal thermoelectric coefficients for a symmetric setup in the clean and dirty limit for various spin mixing parameters $\delta \varphi $ and $G^{\phi}$. 
Both coefficients are normalized to the normal state value of the nonlocal conductance $(L^{qV}_{12})_{T>T_{\rm c}}$, and are plotted in units of $k_{\rm B}T_{\rm c}/|q|$. 
Here, $\PP_n\equiv \PP=0.1$, 
$\eta_\textrm{Th}=\Delta_0$, and
all other parameters are the same as in Fig.~(\ref{phasendiagramm})
}
\end{figure}

We determine the coefficient matrix $\hat{L}$ in Eq.~\eqref{condmat} for temperatures across $T_{\rm c}$. 
We concentrate on the parameters $L^{qT}_{11}$ and $L^{qT}_{12}$, as they are representative for local and nonlocal thermoelectric properties. In Fig.~\ref{tempdep} we plot these parameters for different spin-mixing angles and $10\%$ polarization. 
Remarkably, we obtain qualitatively comparable behaviors of both limits although they are based on very different assumptions. 
The quantitative differences are related to the different shifting mechanisms of the subgap peaks already pointed out above. Hence, the best comparison is found for
small values of $\delta\varphi$ (ballistic) and $G^{\phi}$ (diffusive). We find a zero crossing at a finite temperature in both cases. 
The similarity of local and nonlocal parameters for small temperatures can be understand from the thermally insulating behavior of superconductors at small temperatures. 

We observe that the coefficients in Eq.~\eqref{condmat} fulfill a generalized Onsager symmetry. Onsager's symmetry for local currents was originally derived from microscopic reversibility \cite{onsager}. 
Generalizations of Onsager's reciprocity theorem have been recently discussed
using statistical arguments \cite{utsumi1,utsumi2,utsumi3}.
Here we find a generalization for nonlocal superconductor/ferromagnet three-terminal devices, that include supercurrents as well as crossed Andreev reflection processes.
This follows directly from the analytical formulas \eqref{clean1}-\eqref{clean2} in the clean limit using relations like $\Gamma_j(\varepsilon;-\delta\varphi)=-\Gamma_j^*(-\varepsilon;\delta\varphi)$ (an example is given in the 
Supplementary Material),
and is verified numerically also for the diffusive case. 
This Onsager symmetry holds for any relative angle between the magnetization axes of the two ferromagnets.

In conclusion, we have opened a way of utilizing thermoelectric effects in superconducting spintronics.
This possibility of controlling energy flow in superconducting heterostructures
with spin polarized electrodes allows of a multitude of novel applications.
Particularly interesting for applications is our finding of a zero crossing
in the Seebeck coefficients as function of
temperature, spin polarization, and the relative angle of the magnetization axes. 
This not only would give a possibility to measure spin-filtering parameters and the 
experimentally so far inaccessible spin-mixing parameters, but 
would also allow for sensitive and controllable thermal elements in superconducting circuits.

WB and PM acknowledge financial support from the DFG and the Baden-W\"urttemberg-Stiftung. ME acknowledges support from the EPSRC under grant reference EP/J010618/1. ME and WB were supported from the Excellence Initiative program ``Freir\"aume f\"ur Kreativit{\"a}t'' at the University of Konstanz.

\bibliographystyle{phaip}

\end{document}


\title{Supplementary material for: Nonlocal thermoelectric effects and nonlocal Onsager relations in a three-terminal proximity system}
\author{P. Machon}
\affiliation{Department of Physics, University of Konstanz, D-78457 Konstanz, Germany}
\author{M. Eschrig}
\affiliation{Department of Physics, Royal Holloway, University of London, Egham, Surrey TW20 0EX, UK}
\author{W. Belzig}
\affiliation{Department of Physics, University of Konstanz, D-78457 Konstanz, Germany}
\maketitle

\section{Spin dependent quasiclassical theory in the dirty limit}
In the following we will give a more detailed description of our calculations. In the stationary case the non-equilibrium Keldysh Green function in Fourier presentation $\check{\cal G}(\rv,\rv',\e)=\int\,dt/\hbar\,\check{\cal G}(\rv,\rv',t-t')\exp{(i\e(t-t')/\hbar)}$ reads:
\begin{equation*}\begin{aligned}
\check{\cal G}(\vec{r},\vec{r}^{\,\prime},\e)=\left(
\begin{array}
[c]{cc}%
\check{\cal G}^{\rm R}(\vec{r},\vec{r}^{\,\prime},\e) & \check{\cal G}^{\rm K}
(\vec{r},\vec{r}^{\,\prime},\e)\\
0 & \check{\cal G}^{\rm A}(\vec{r},\vec{r}^{\,\prime},\e)
\end{array}\right)\end{aligned}
\end{equation*}
with
{\small{
\begin{align*}
&\check{\cal G}^{\rm R/A}(\vec{r},\vec{r}^{\,\prime},t-t^{\prime})=\mp i\theta
(\pm(t-t^{\prime}))\left\langle \left\{  \hat{\Psi}
(t,\vec{r}),\hat{\Psi}^{\dagger}(t^{\prime},\vec{r}^{\,\prime})\right\}
\right\rangle\\ 
&\check{\cal G}^{\rm K}(\vec{r},\vec{r}^{\,\prime},t-t^{\prime})=-i\left\langle \left[  \hat{\Psi}(t,\vec{r}),\hat{\Psi}^{\dagger
}(t^{\prime},\vec{r}^{\,\prime})\right]\right\rangle.
\end{align*}}}
Here ${\rm R/A/K}$ labels the retarded, advanced, and Keldysh part respectively.
We choose the spinor basis 
\begin{align*}
\hat{\Psi}\dg(t,\rv)=\left(\Psi\dg\ua(t,\rv),\Psi\dg\da(t,\rv),\Psi\da(t,\rv),-\Psi\ua(t,\rv)\right).
\end{align*}
In the following, $\vec{\check\sigma}$ and $\check\sigma_0$ denote the vector of Pauli matrices and the unit matrix in spin space, respectively, and $\check\tau_z$ denotes the third Pauli matrix in Nambu space. 
In this basis the quasiclassical isotropic Green functions of a bulk BCS-superconductor (S) and a ferromagnet (F) are given by ($\delta>0$, $\delta\rightarrow0$)
\begin{align*}
\begin{aligned}\G_{S}^{\rm R/A}&=\frac{\pm\sign(\e)}{\sqrt{(\e\pm i\delta)-|\Delta|^2}}\begin{pmatrix}(\e\pm i\delta)&\Delta^*\\-\Delta&-(\e\pm i\delta)\end{pmatrix}\otimes\check\sigma_0\\\G_F^{\rm R/A}&=\pm\check\tau_z\otimes\check\sigma_0.\end{aligned}
\end{align*}
Here, $\Delta$ is the superconducting order parameter.
Note that the definition of $\G_F$ is equal to the one of a normal metal. Ferromagnetism enters only in the SDBC via spin-dependent phase shifts and interface polarization effects.
According to, e.g., review \cite{belzig} the Keldysh component is 
\begin{align*}
&\G^{\rm K}=\G^{\rm R}\check{h}-\check{h}\G^{\rm A},
\end{align*}
with the distribution matrix
\begin{align*}
&\check{h}=\begin{pmatrix}\tanh{\frac{\e-qV}{2T}}&0\\0&\tanh{\frac{\e+qV}{2T}}\end{pmatrix}\otimes\check\sigma_0.
\end{align*}
Following the steps done in Ref. \cite{cottet} the SDBC in the tunneling limit 
follow only for the limit of small ${\cal P}_n$. Thus, we derived a version of the boundary conditions following \cite{eschrigkopu}, which is valid for general ${\cal P}_n$. In linear order in ${\cal T}_n$ and $\delta \varphi_n$ we obtain Eq.~(3) of the Letter, which is written explicitely as
[$\check{\kappa}=\check\tau_z\otimes(\vec{m}\vec{\check\sigma})$, where
the ferromagnet is described by its magnetization direction unit vector $\vec{m}$]
\begin{align*}
\begin{aligned}
2\check{\cal I}_{L}&(\varepsilon)=
2G_q\mbox{$\sum_n$} \left[\check t_n^{LR} \check{\cal G}_{R}(\varepsilon) (\check t_n^{LR})^\dagger - i\delta \varphi^{L}_{n} \check \kappa, \check{\cal G}_{L}(\varepsilon )\right]
\\
=&G^0\left[\check{\cal G}_{R}(\varepsilon),\check{\cal G}_{L} (\varepsilon)\right]+
G^{\text{MR}}\left[\left\{\check{\kappa},\check{\cal G}_{R}(\varepsilon)\right\}  ,\check{\cal G}_{L}(\varepsilon)\right]\\
& 
+{G}^{1}\left[\check{\kappa}\check{\cal G}_{R}(\varepsilon)\check{\kappa},\check{\cal G}_{L}(\varepsilon)\right] -iG^{\phi}_{}\left[  \check{\kappa},\check{\cal G}_{L}(\varepsilon)\right] 
\end{aligned}
\end{align*}
($G_q=q^2/h$ is the conductance quantum) with
\begin{eqnarray*}
G^{0}  &=& G_q
\mbox{$\sum\nolimits_{n}$}
{\cal T}_{n}\label{GT}\left( 1 + \sqrt{1-{\cal P}_n^2}\right)\\
{G}^{1}  &  =&G_q
\mbox{$\sum\nolimits_{n}$}
{\cal T}_{n}\left( 1- \sqrt{1-{\cal P}_{n}^2}\right)\label{GMR2}\\
G^{\text{MR}}    &=&G_q
\mbox{$\sum\nolimits_{n}$}
{\cal T}_{n}{\cal P}_{n}\label{GMR} \\
G^{\phi}_{}    &=&2G_q
\mbox{$\sum\nolimits_{n}$}
\delta \varphi_{n}^L\label{Gfi} 
\end{eqnarray*}
$\G_{R/L}$ are the Green functions of the right ($R$) and left ($L$) side respectively. 
Note that as $\G_F^{\rm R/A/K}$ commutes with $\hat \kappa$, the terms with $G^0$ and ${G}^{1}$ can in our case be combined to one term of the form $G\left[\check{\cal G}_{F},\check{\cal G}_{c}\right]$ with $G=G^0+{G}^{1}=
2G_q\sum\nolimits_{n}{\cal T}_{n}$, leading to a simplified boundary condition for $F/S$ contact $j$ [for notational simplicity we use for each $F$ terminal the same index as for the corresponding contact, $j$, e.g. we write $\check {\cal G}_j(\varepsilon)\equiv \check {\cal G}_F(V_j, T_j; \varepsilon)$]
\begin{align*}
\begin{aligned}
\check{\cal I}_{j, c}(\varepsilon)=&\frac{1}{2}\left[G_j\check{\cal G}_{j}
(\varepsilon)
+ G_j^{\text{MR}}\left\{\check{\kappa}_{j},\check{\cal G}_{j}
(\varepsilon)
\right\}
-iG^{\phi}_{j}\check{\kappa}_{j},\check{\cal G}_{c}
(\varepsilon)
\right].
\end{aligned}
\end{align*}
For the contact to the bulk superconductor the boundary condition is
$\check{\cal I}_{S ,c}(\varepsilon)=\frac{1}{2}G_{S}\left[\check{\cal G}_{S}(\varepsilon),\check{\cal G}_{c}(\varepsilon)\right]$.
Dephasing in the contact region is described by a leakage terminal,
$\check{\cal I}_{\rm Leak}(\varepsilon)=(G_q/4\varepsilon_{\rm Th}) 
[\G_{\rm Leak}(\varepsilon),\G_{c}(\varepsilon)]$, 
with
$\G_{\rm Leak}^{\rm R/A}(\varepsilon)=-i(\e\pm i\delta)\check\tau_z\otimes\check\sigma_0$
and 
Thouless energy $\e_{\rm Th}$. 
The Green function of the contact region, $\G_c$, fulfills a Kirchhoff rule \cite{nazarov}
$\check{\cal I}_{\rm Leak}+\check{\cal I}_{S ,c}+\sum\nolimits_j\check{\cal I}_{j,c}=0$,
and hence is
is determined from equations of the form $[\check{\cal M},\check{\cal G}_c]=0$ with normalization condition $\check{\cal G}_c^2=\check 1$. 
The charge and energy currents are obtained from the Keldysh component of the matrix current \cite{morten}:
\begin{subequations}\label{current:dirty}
\begin{align*}
I^q_{j}&=\frac{1}{8q}\int\text{Tr}\left[
(\check{\tau}_z\otimes\check{\sigma}_0)
\check{\cal I}_{j,c}^{\scriptstyle{\rm K}}(\varepsilon) 
\right] d\varepsilon\\
I^{\epsilon}_{j}&=\frac{1}{8q^2}\int \e\text{Tr}\left[\check{\cal I}_{j,{\rm  c}}^{\scriptstyle{\rm K}}(\varepsilon)\right]d\varepsilon.
\end{align*}
\end{subequations}
The density of states in the contact region is defined from its retarded Green function ${\cal G}^{\rm R}_c$ like 
\begin{align*}
D(\varepsilon)=\frac{1}{4}\text{Tr}[\check\tau_z\otimes\check\sigma_0\check{\cal G}^{\rm R}_c(\varepsilon) ] .
\end{align*}

\begin{widetext}
\section{Classification of local and nonlocal processes for ballistic three-terminal systems}

In this section we show examples for transport processes that contribute to the various terms of Eqs. (6)-(9) of the Letter. These terms involve propagation of particles or holes, which will be represented as full lines and dashed lines in the figures following. Certain processes involve a conversion between particles and holes, accompanied by the creation or destruction of a Cooper pair. These processes will be represented by a loop, where a full line turns into a dashed line or vice versa. These loops in the following figures correspond to the factors $|\gamma_0|^2$ in Eqs. (6)-(9) of the Letter. Propagation between the left and right interface is represented in these equations by the factors $u_{12}$ and $u_{21}$. Vertex corrections $v_1$ and $v_2$ correspond to multiple Andreev reflections possible at both interfaces. The factors $\gamma_1$ and $\gamma_2$ 
combine propagation between the two interfaces with Andreev reflections at the other interface.
\begin{figure*}[h]
\includegraphics*[width=0.2\linewidth,clip]{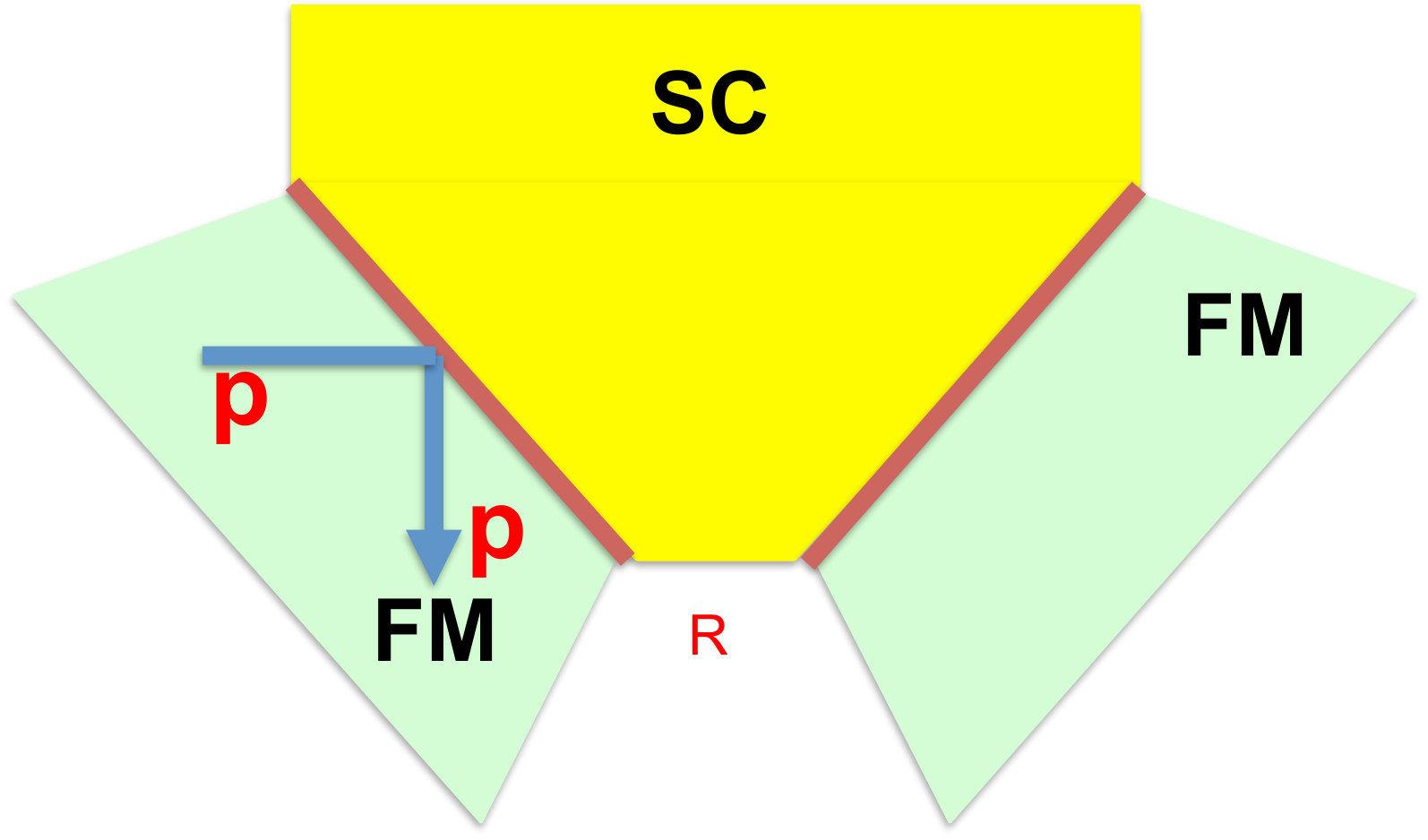}
\includegraphics*[width=0.2\linewidth,clip]{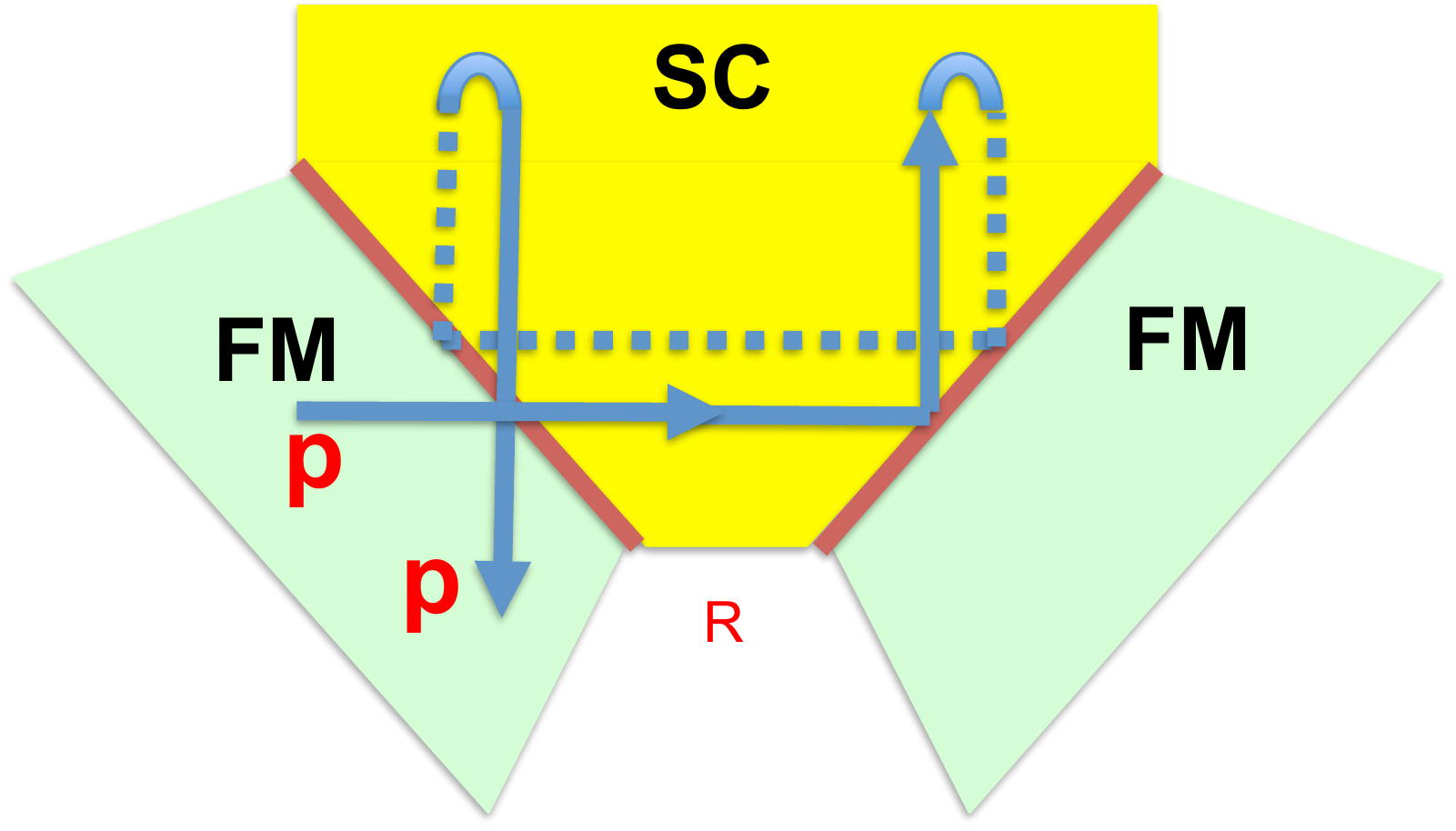}
\includegraphics*[width=0.2\linewidth,clip]{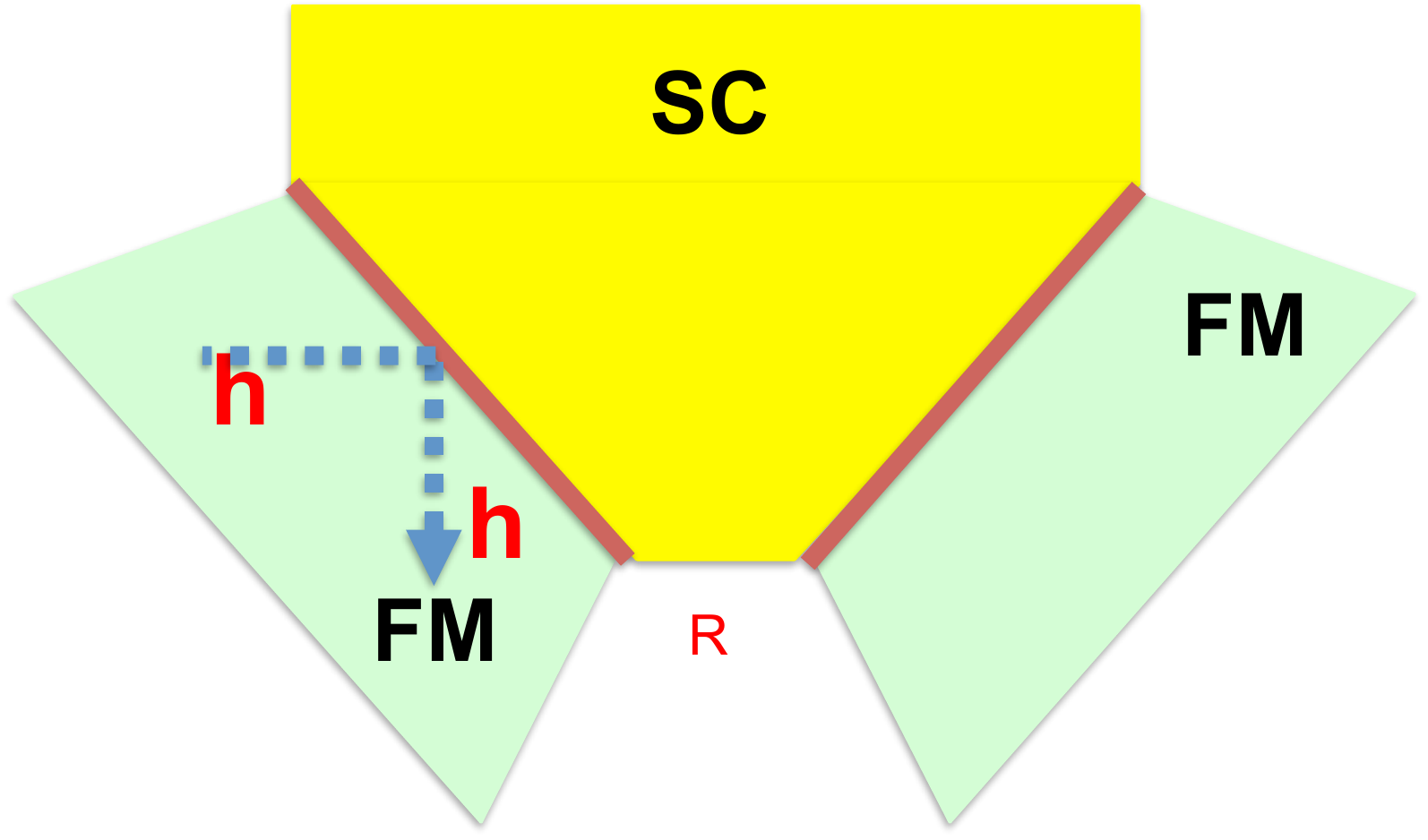}
\includegraphics*[width=0.2\linewidth,clip]{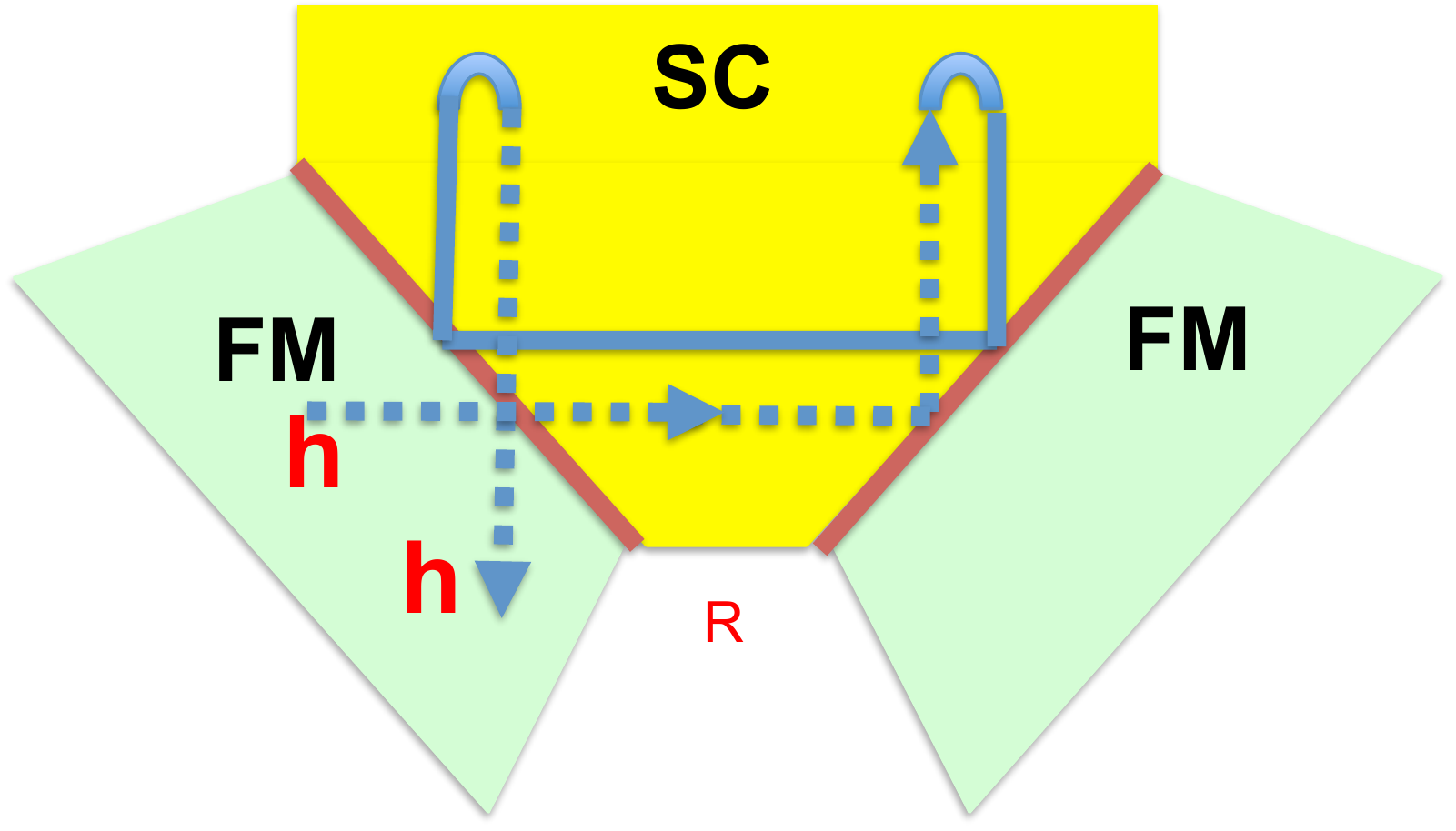}
\\
\includegraphics*[width=0.2\linewidth,clip]{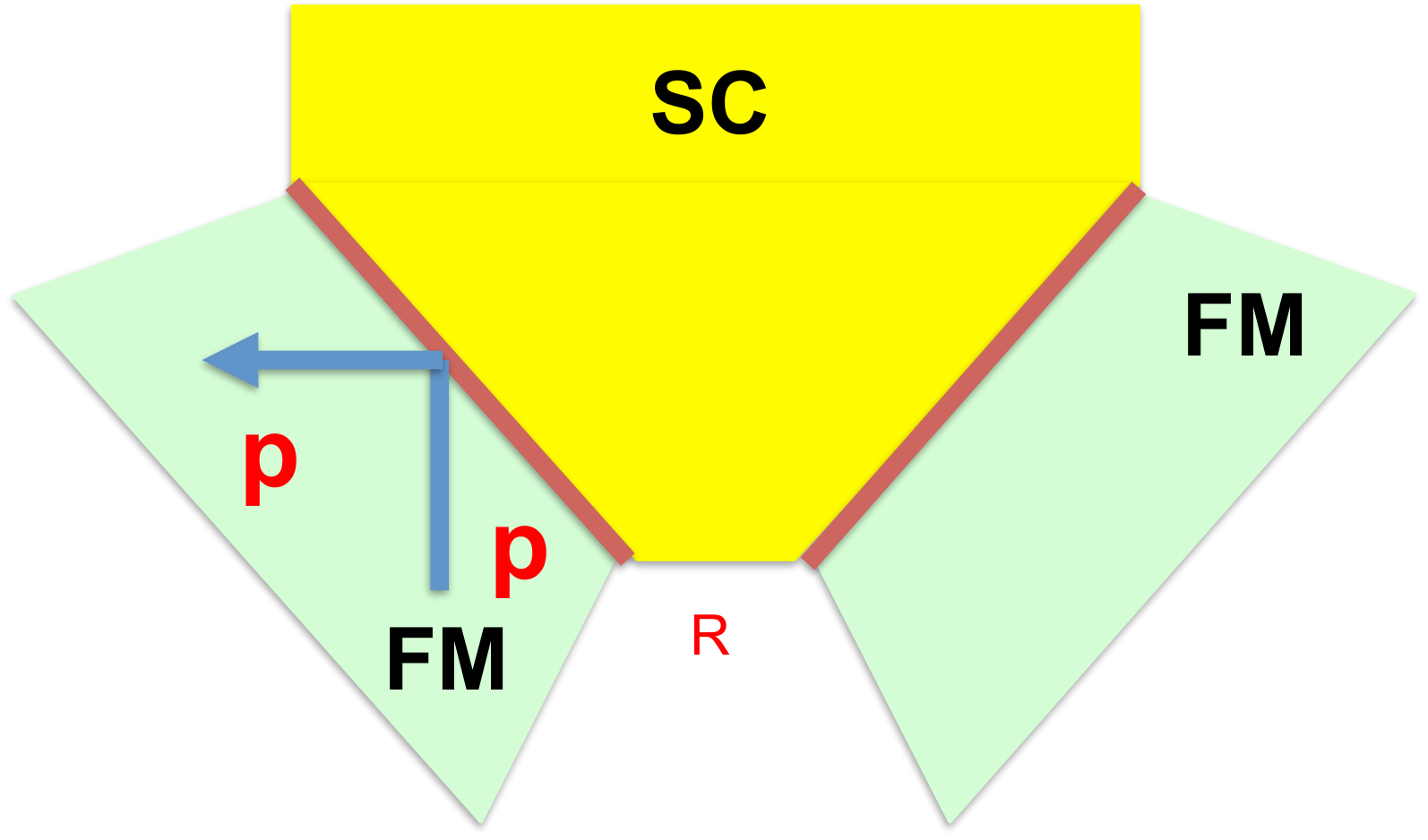}
\includegraphics*[width=0.2\linewidth,clip]{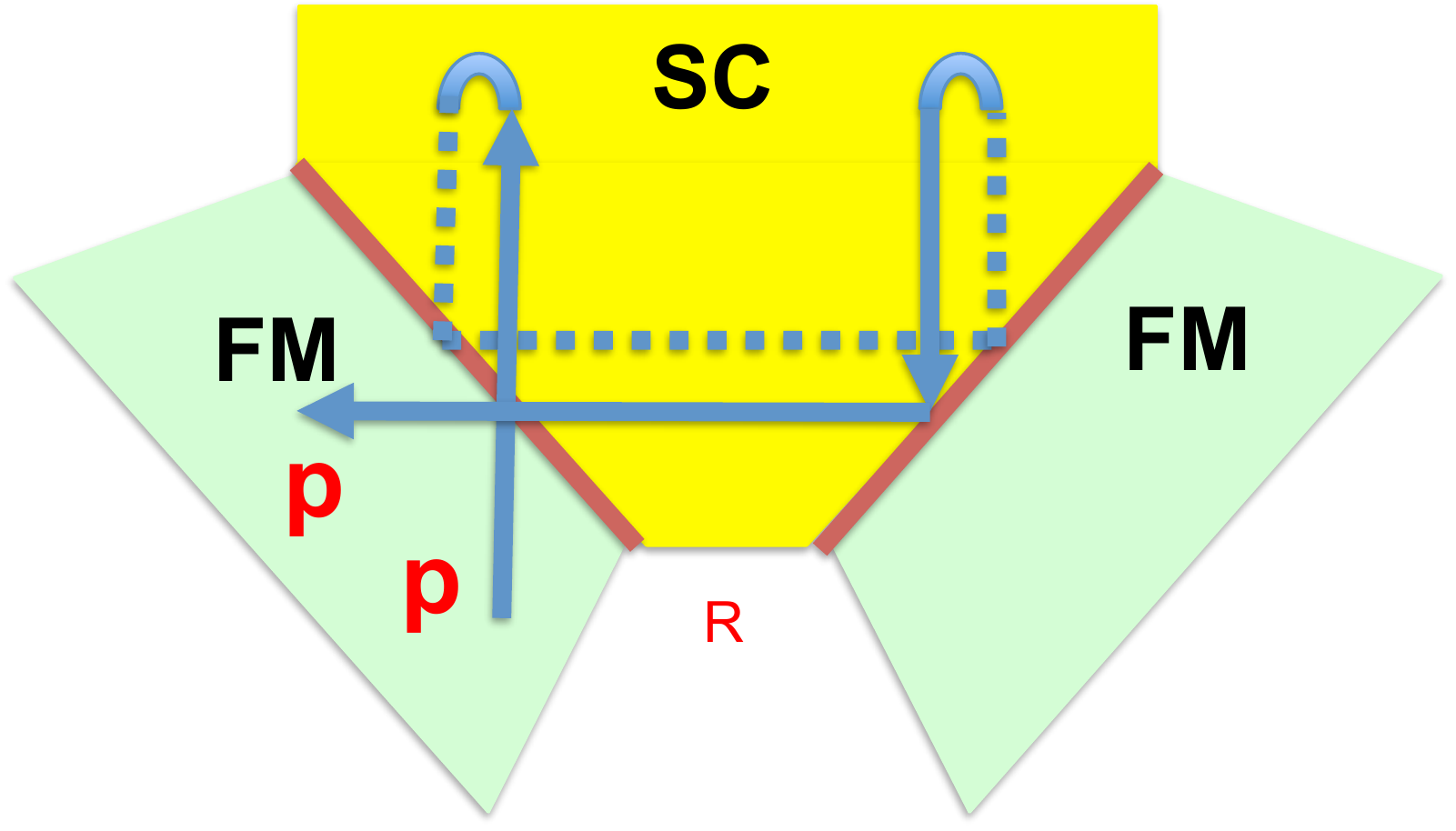}
\includegraphics*[width=0.2\linewidth,clip]{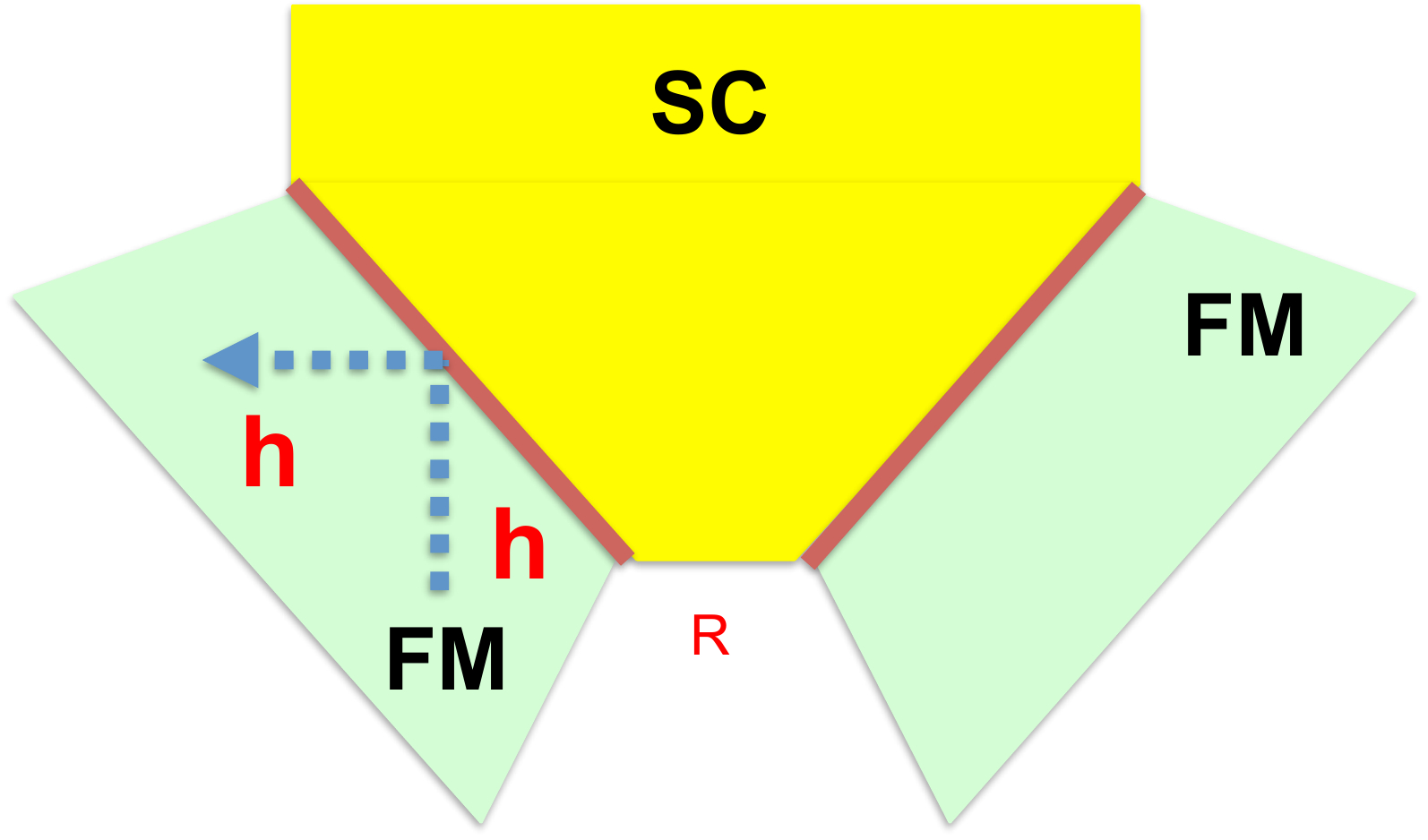}
\includegraphics*[width=0.2\linewidth,clip]{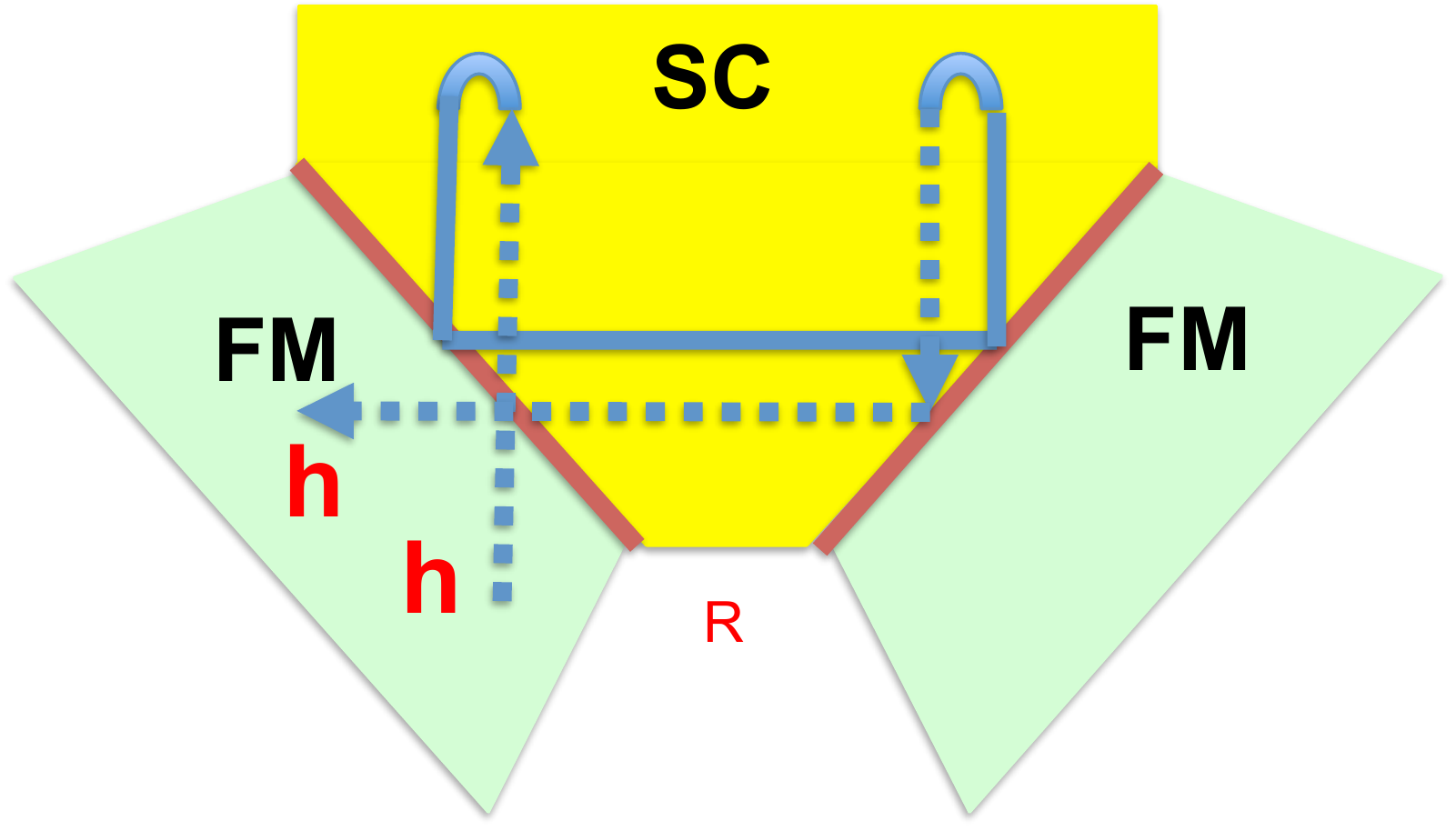}
\\
\caption{\label{IR}
Various contributions to the reflection components $I_{1,\rm R}$.
These processes are characterized by effectively scattering a particle into a particle or a hole into a hole at the same interface.
Only the leading terms are shown, with up to two  Andreev reflections [denoted as the loops turning particles (full lines) into holes (dashed lines) or vice versa].
}
\includegraphics*[width=0.2\linewidth,clip]{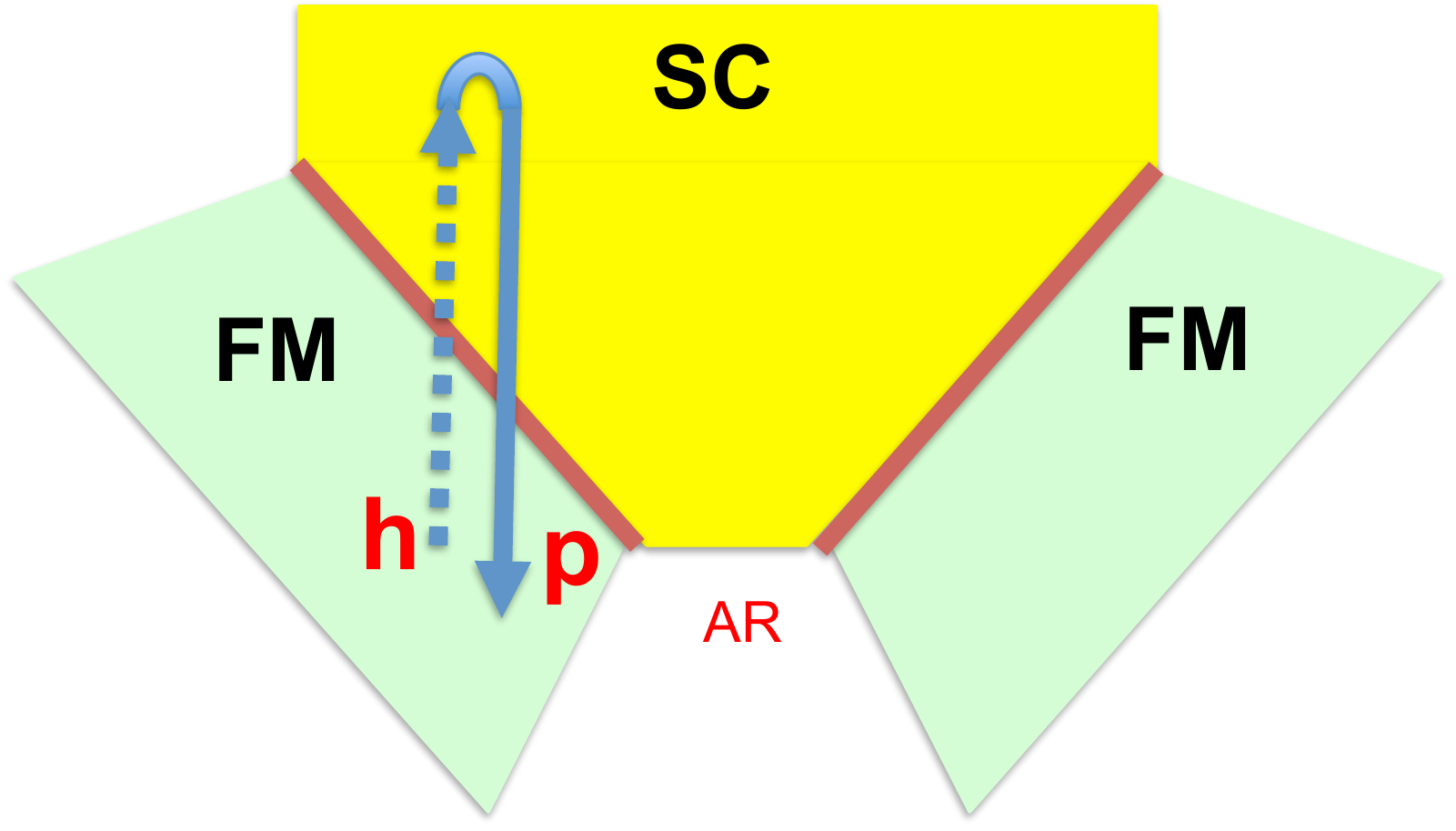}
\includegraphics*[width=0.2\linewidth,clip]{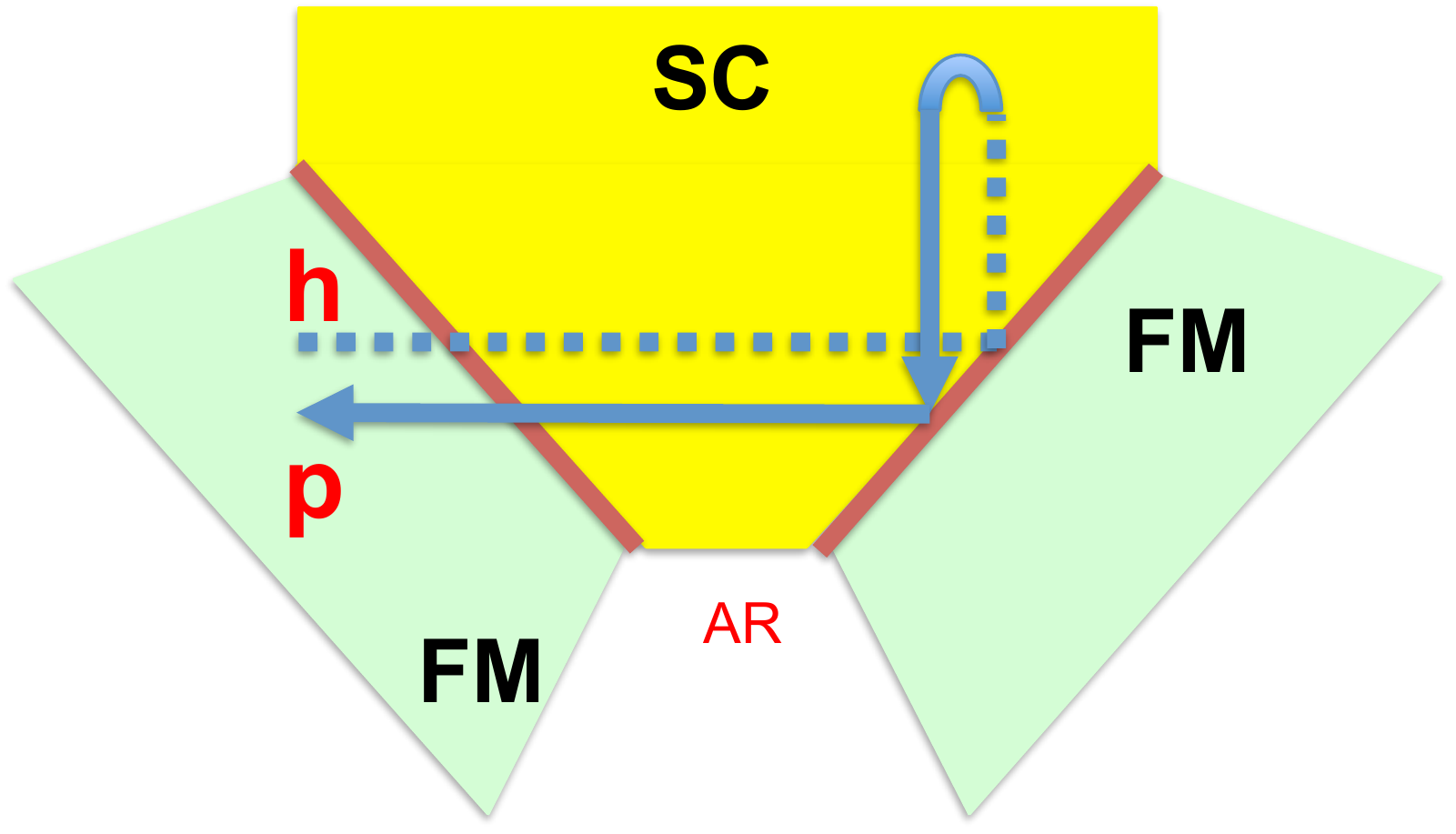}
\includegraphics*[width=0.2\linewidth,clip]{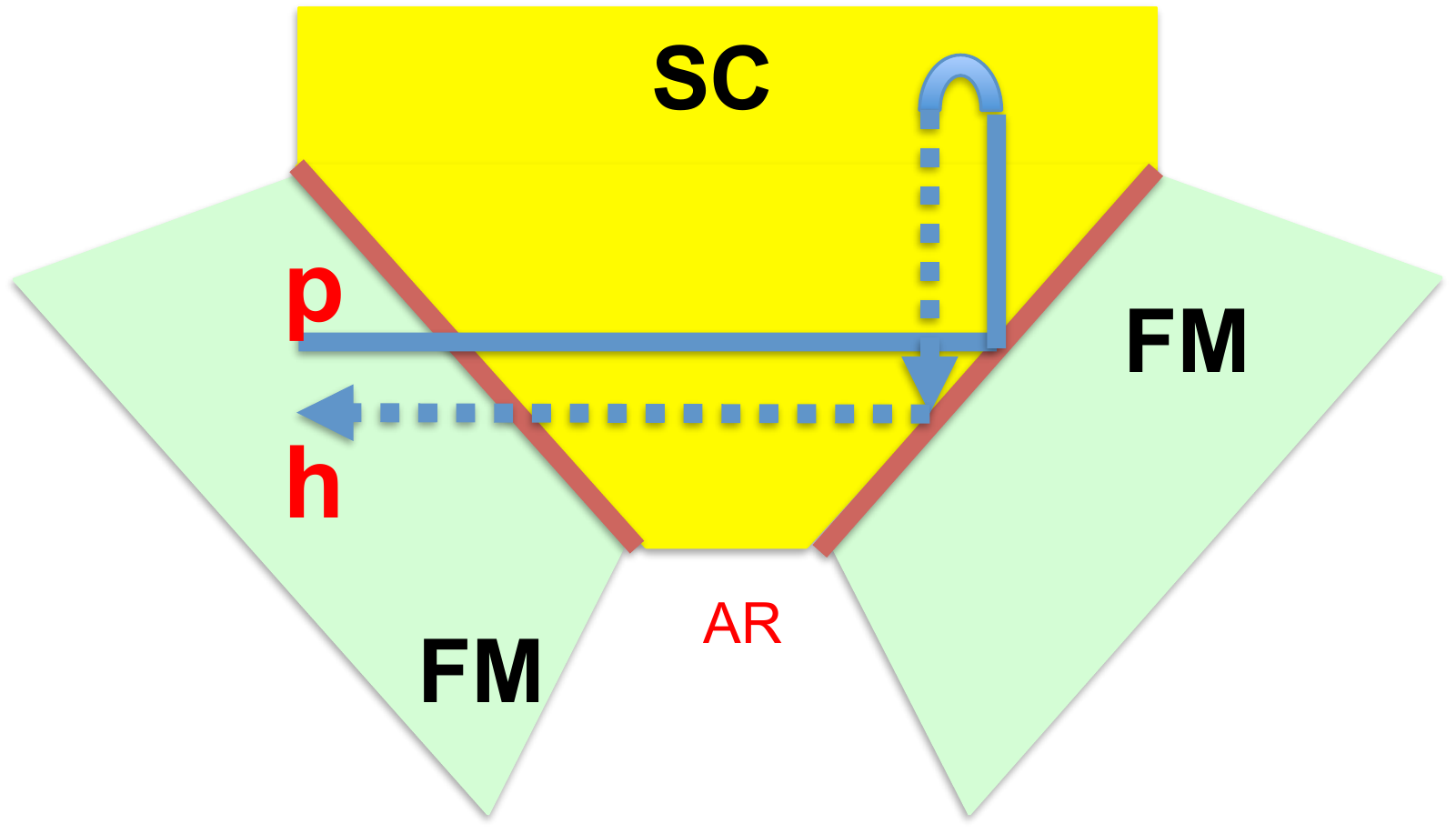}
\includegraphics*[width=0.2\linewidth,clip]{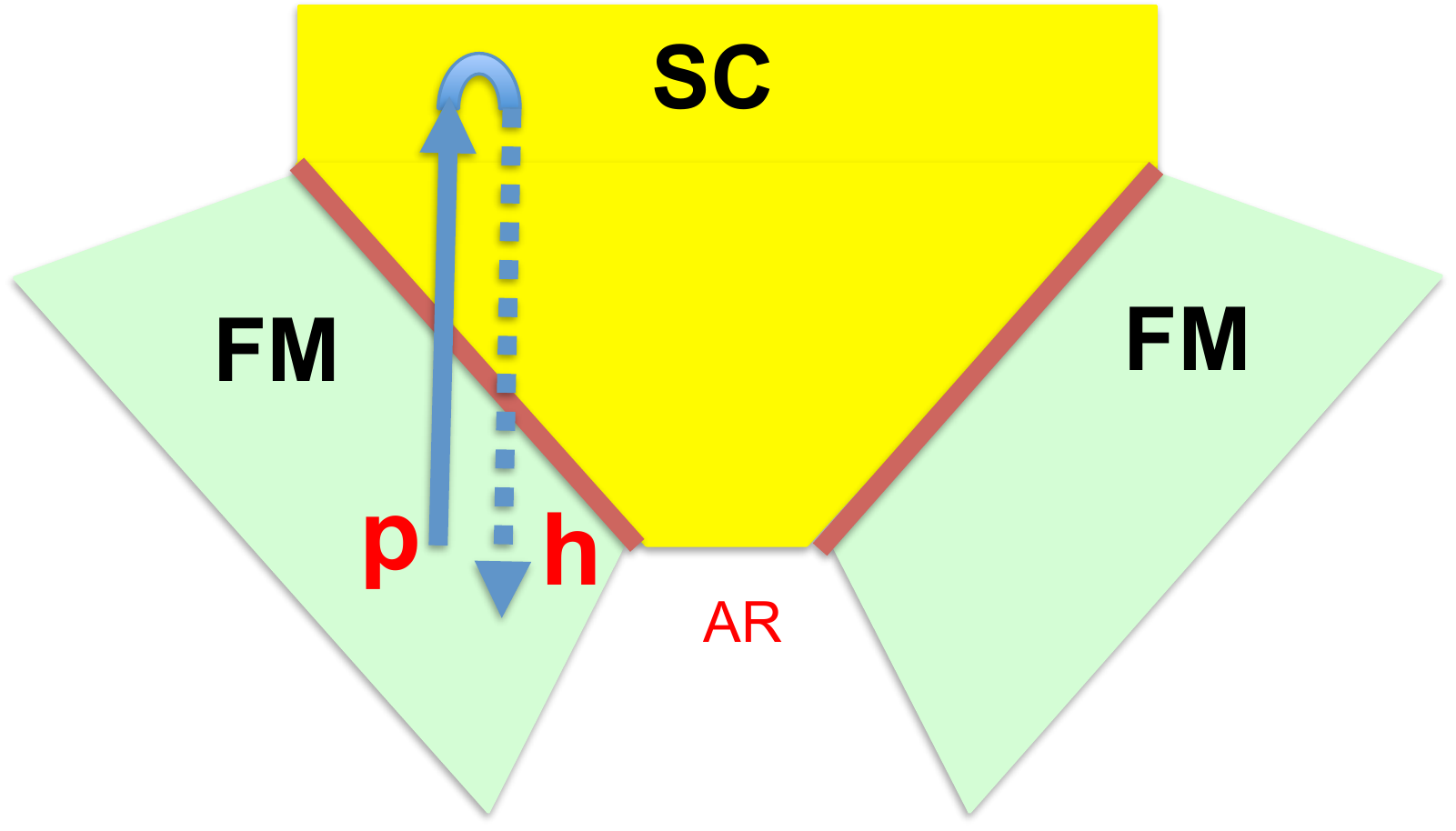}
\caption{\label{IAR}
Various contributions to the Andreev reflection components $I_{1,\rm AR}$. 
These processes are characterized by effectively scattering a particle into a hole or a hole into a particle at the same interface.
Only the leading terms are shown, with one Andreev reflection.
}
\includegraphics*[width=0.2\linewidth,clip]{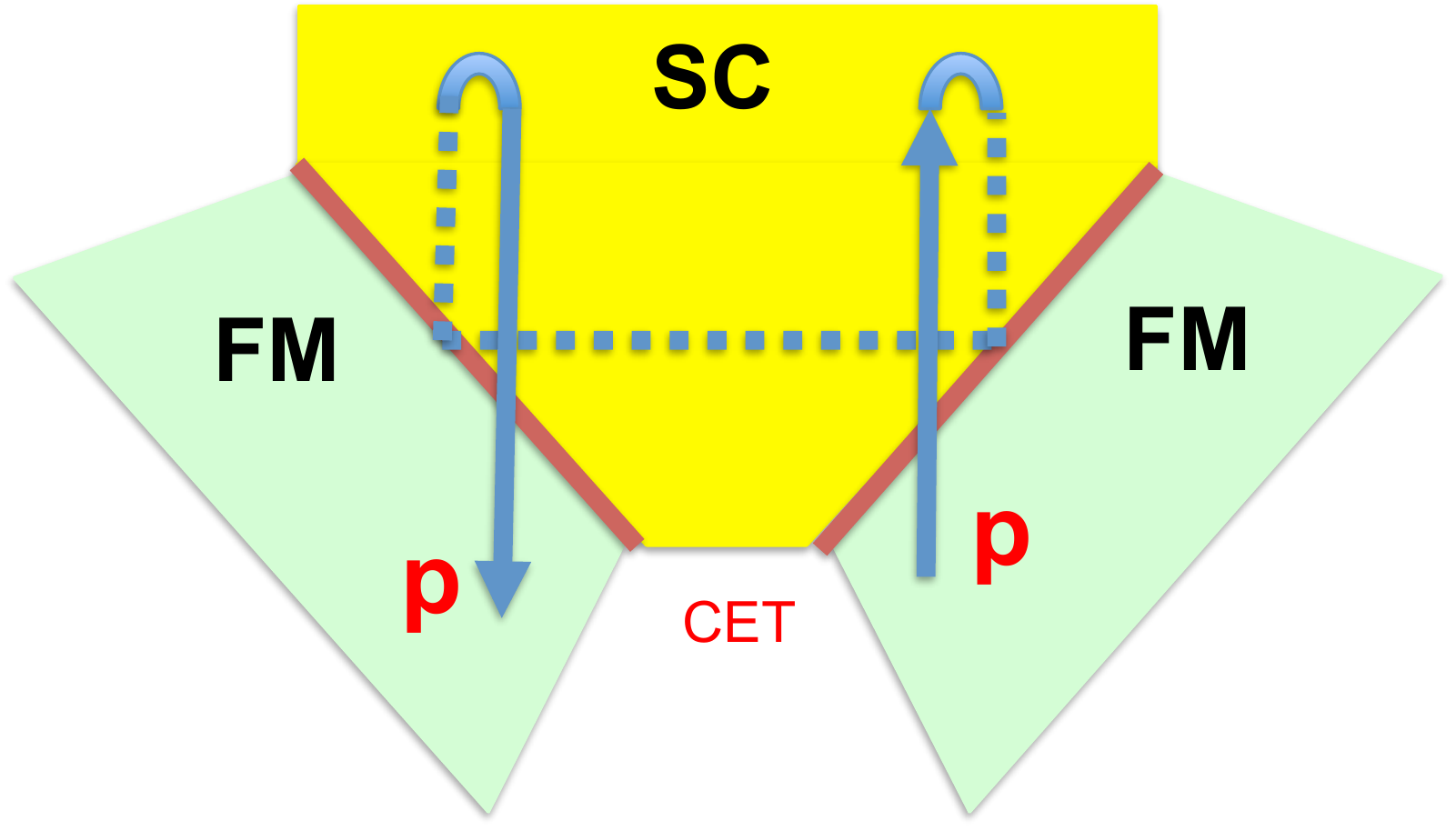}
\includegraphics*[width=0.2\linewidth,clip]{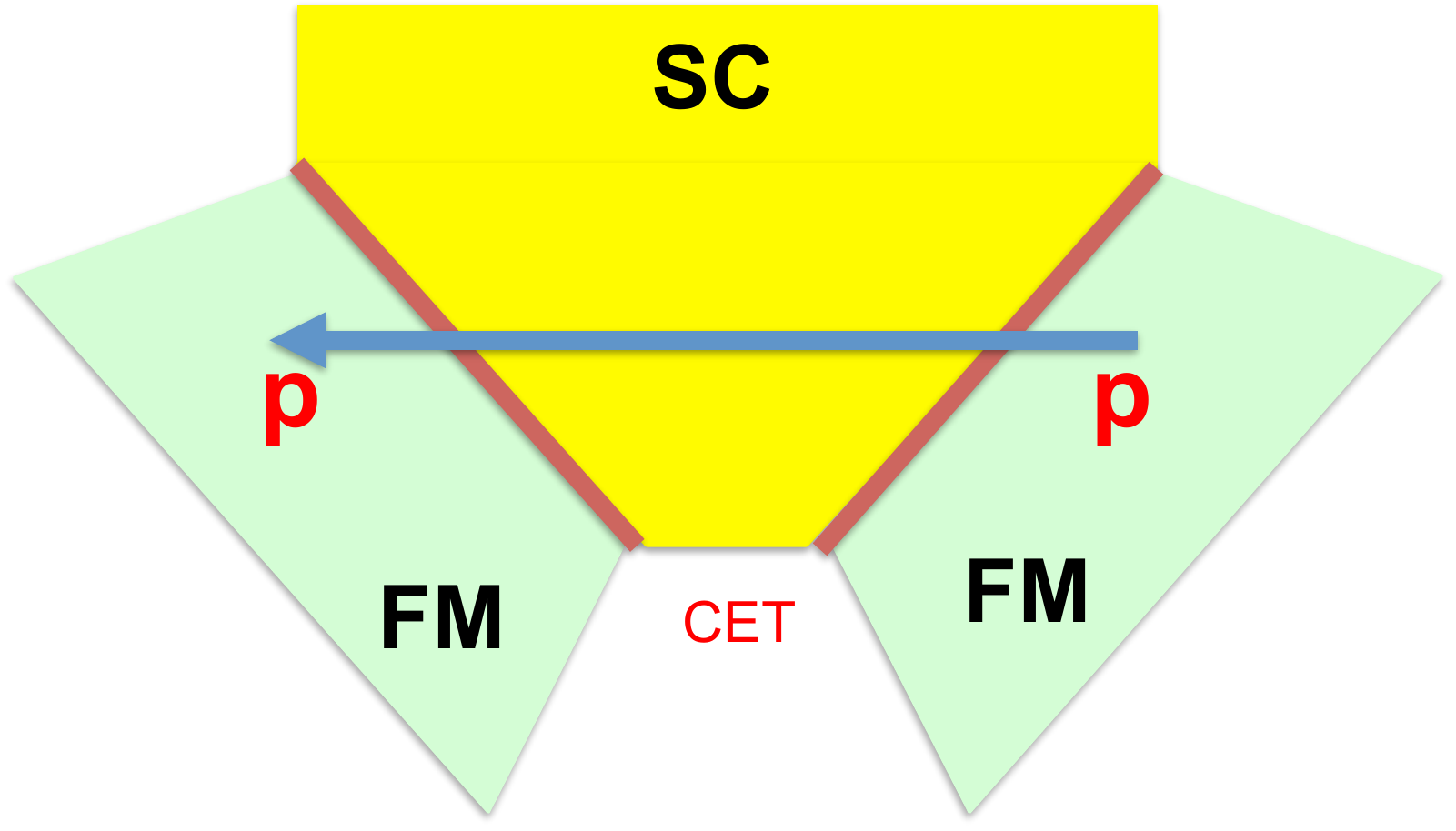}
\includegraphics*[width=0.2\linewidth,clip]{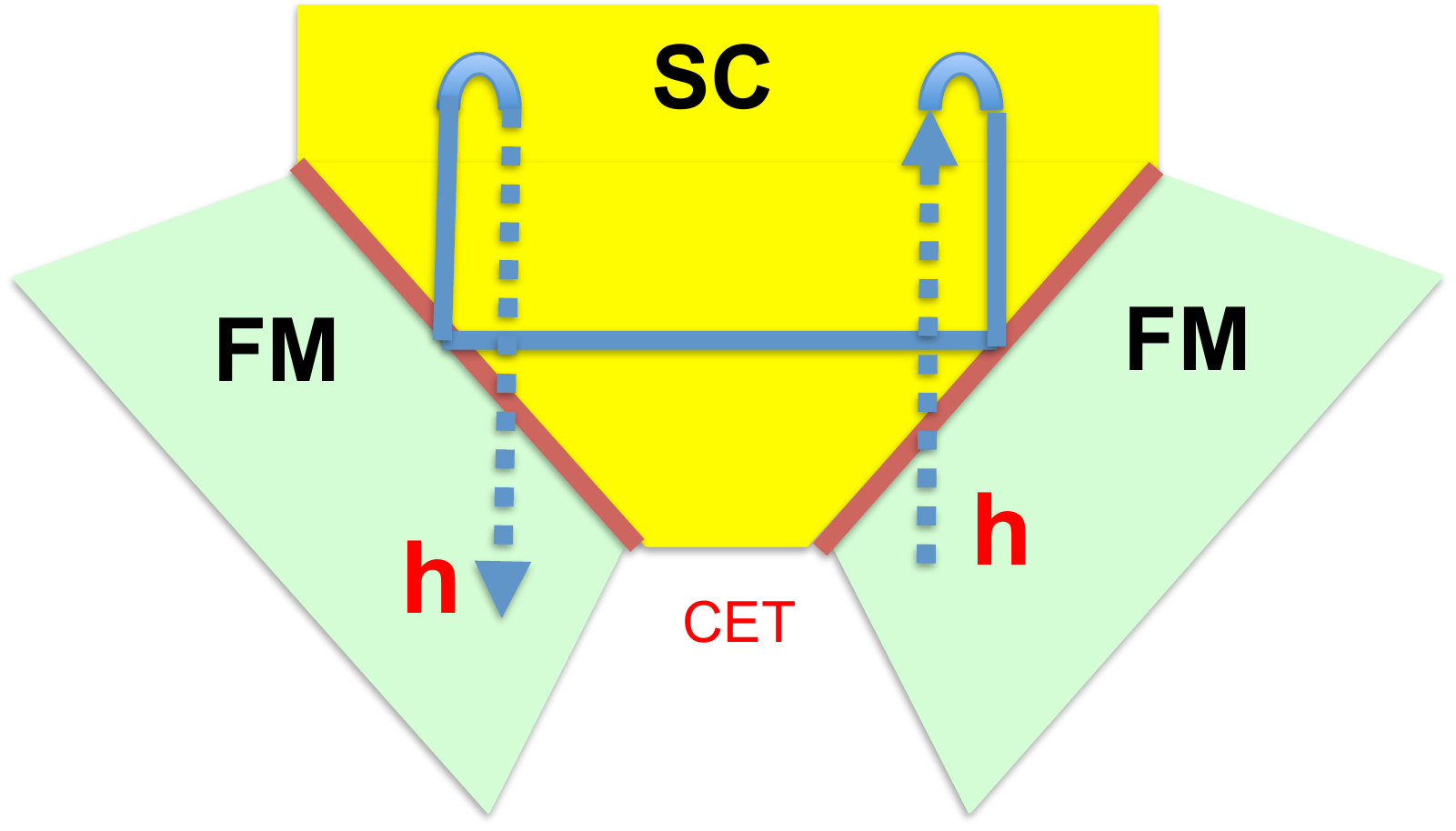}
\includegraphics*[width=0.2\linewidth,clip]{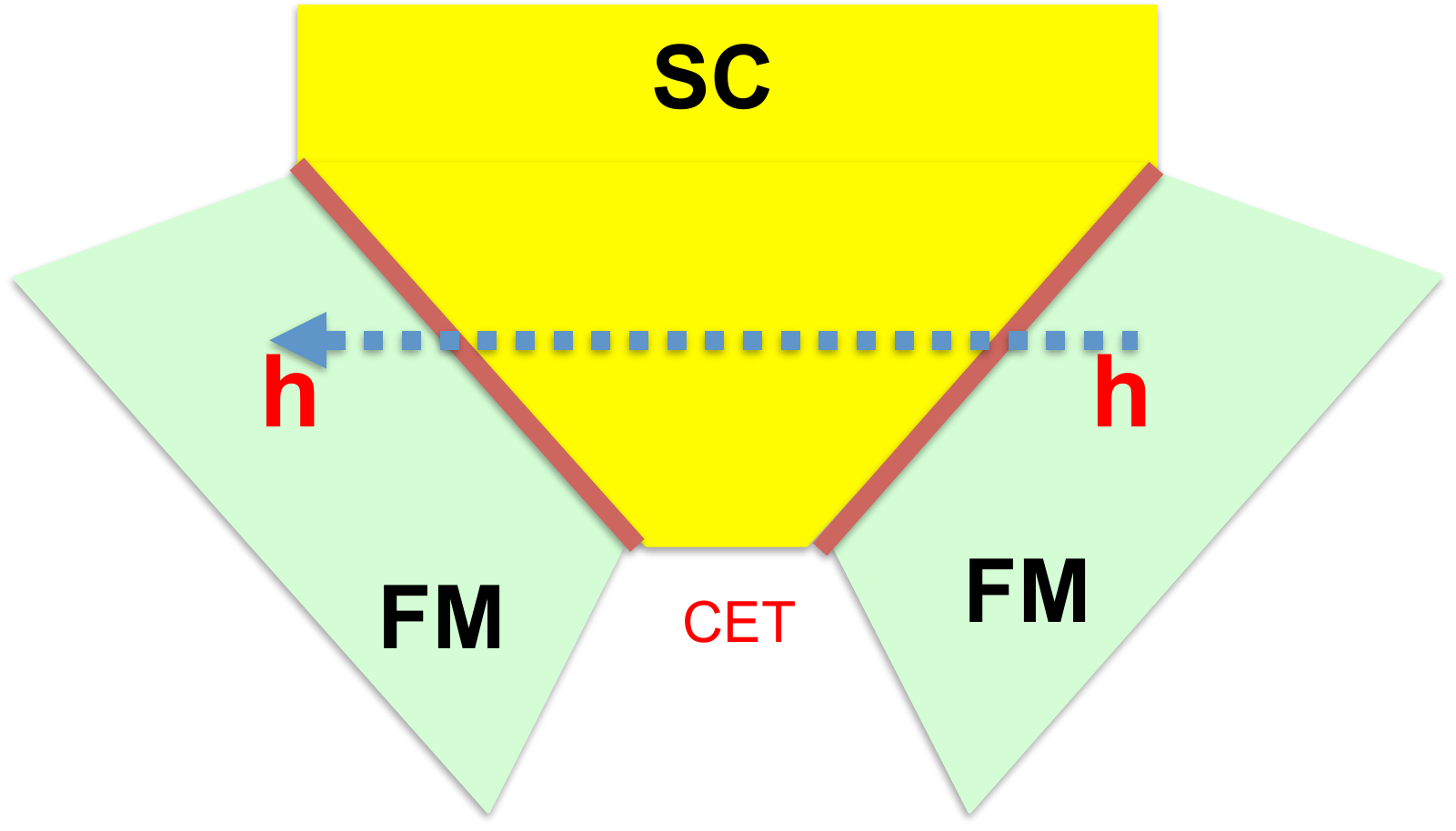}
\caption{\label{ICAR}
Various contributions to the coherent electron transfer components $I_{1,\rm CET}$.
These processes are characterized by effectively scattering a particle from one interface into a particle at the other interface or a hole at one interface into a hole at the other interface.
Only the leading terms are shown, with up to two Andreev reflections.
}
\includegraphics*[width=0.2\linewidth,clip]{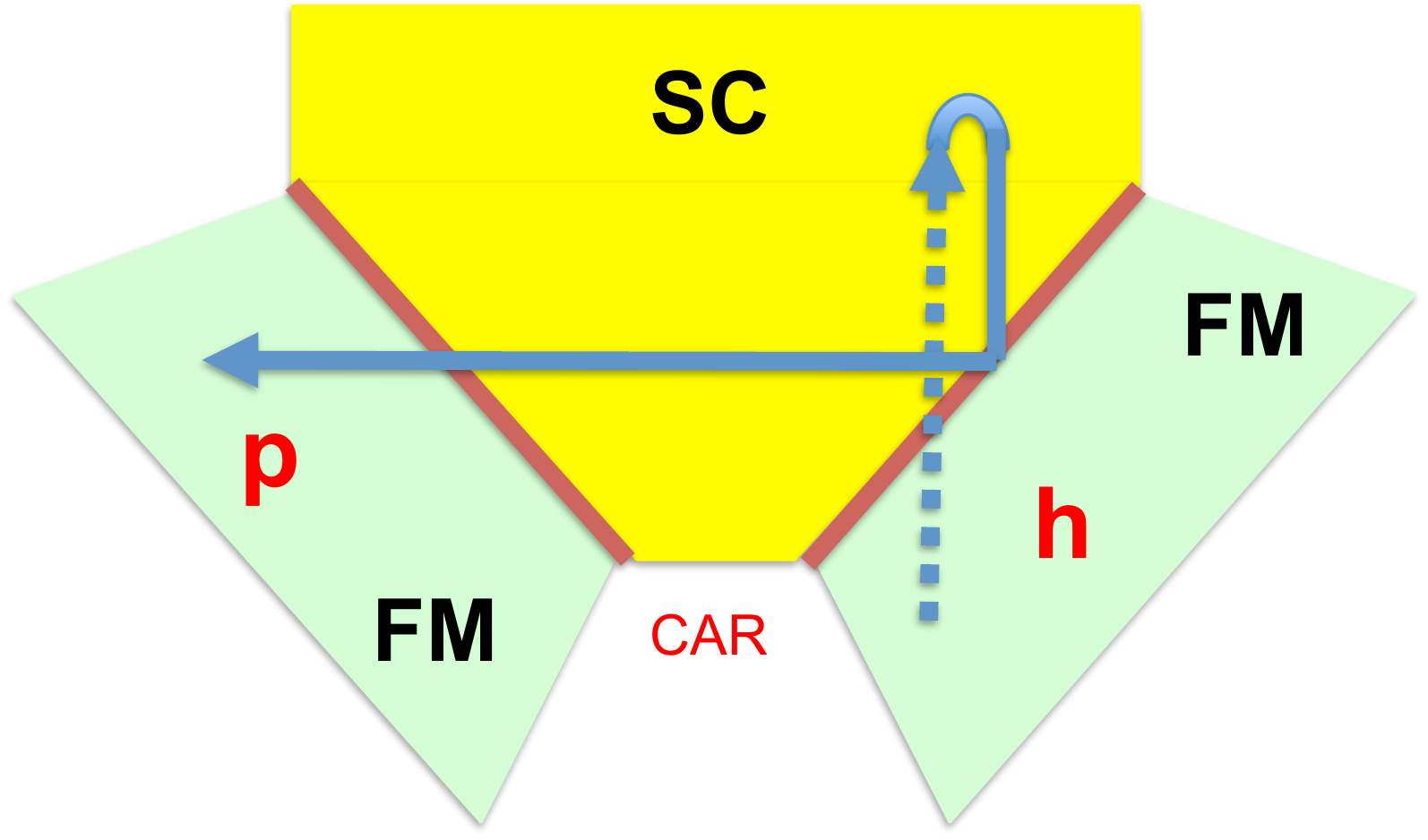}
\includegraphics*[width=0.2\linewidth,clip]{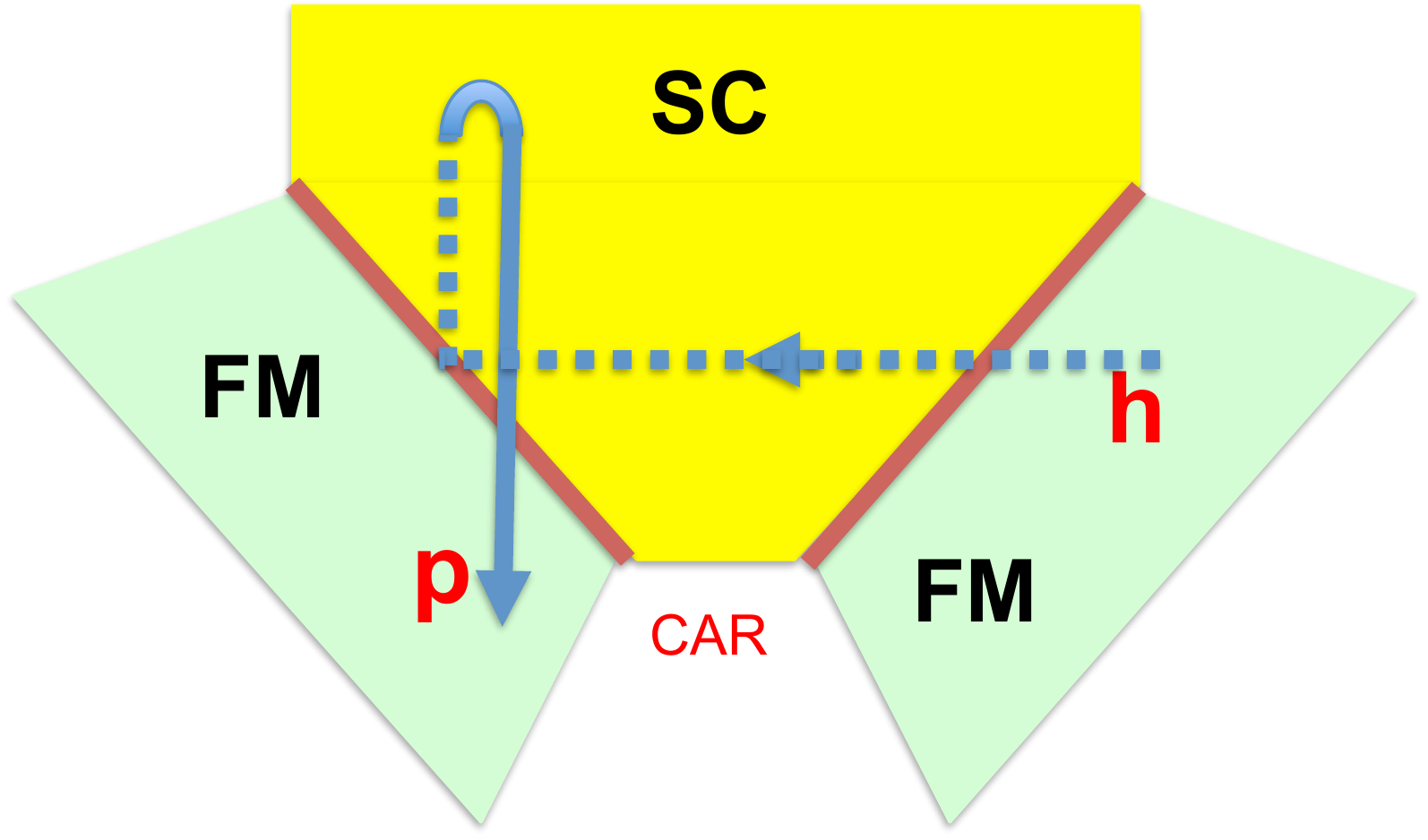}
\includegraphics*[width=0.2\linewidth,clip]{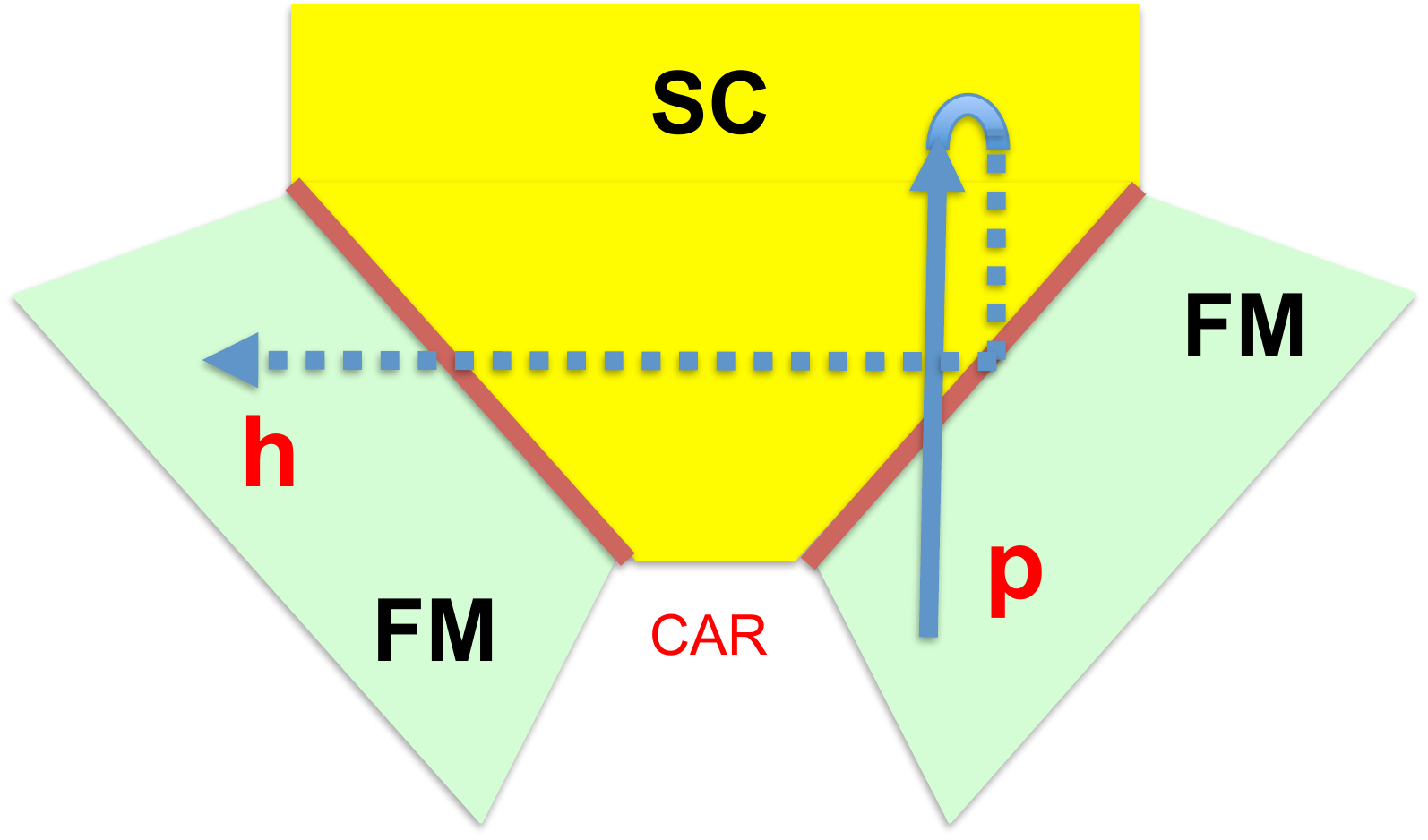}
\includegraphics*[width=0.2\linewidth,clip]{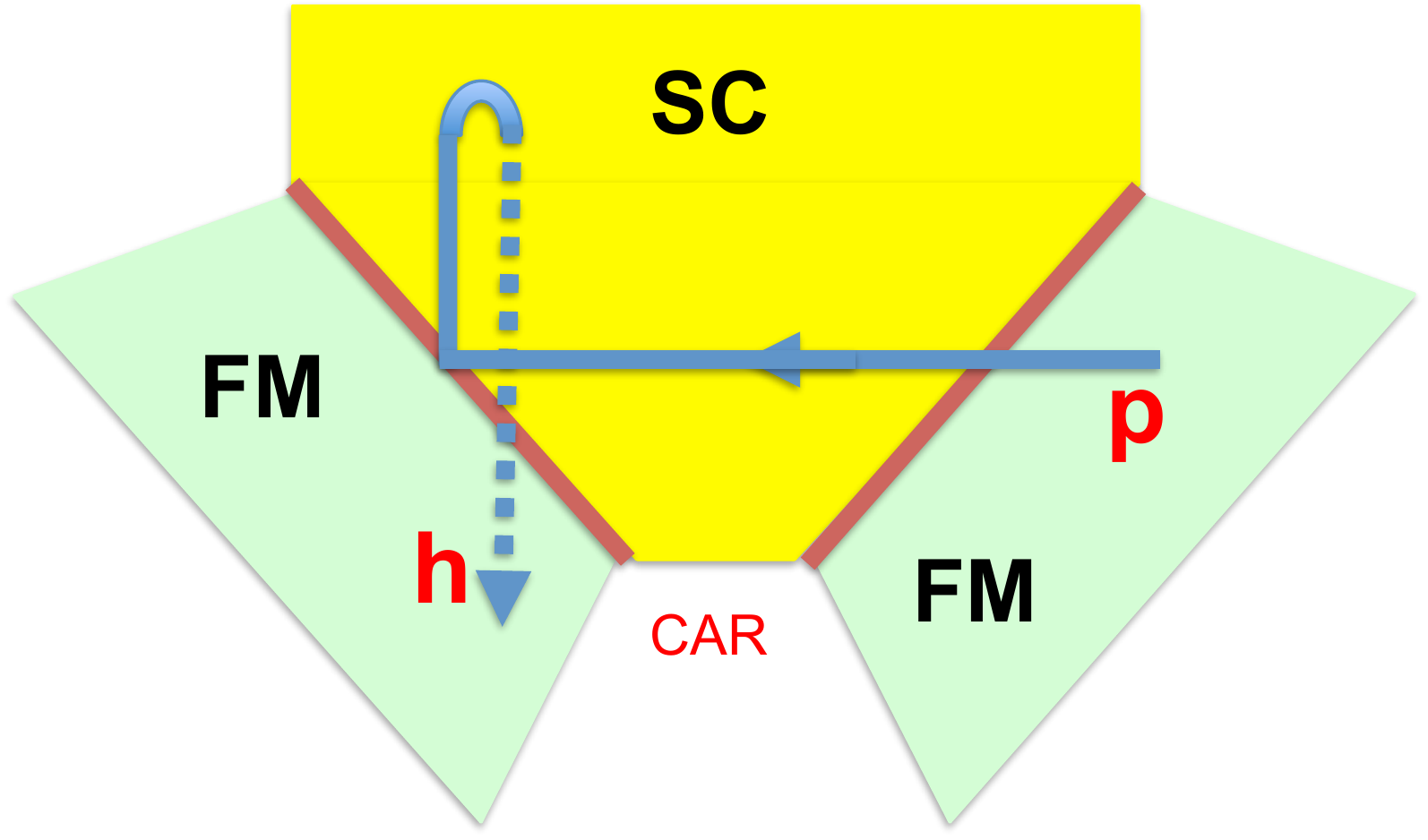}
\caption{\label{ICET}
Various contributions to the crossed Andreev reflection components $I_{1,\rm CAR}$.
These processes are characterized by effectively scattering a particle from one interface into a hole at the other interface or a hole at one interface into a particle at the other interface.
Only the leading terms are shown, with Andreev one reflection.
}
\end{figure*}

We underline that Eqs. (6)-(10) of the Letter were obtained by calculating the full expression for the current through each interface. This expression can then be split into the various terms in a natural way, with the interpretation as shown in the figures above.

\end{widetext}
\newpage
\section{Analytic proof of Onsager's symmetry in the clean limit}
Concerning the Onsager symmetry pointed out in the Letter we will show in the following how the symmetry follows analytically from the formulas (5)-(10) of the Letter in the clean case. As an example we will choose the nonlocal relation $L^{qT}_{12}=L^{\varepsilon V}_{21}$. 
Using 
\begin{align*}
\frac{\partial f_{i,\rm p}}{\partial [\Delta V_j]}&=\delta_{ij} \frac{q}{4k_{\rm B}T_{S} \cosh^2(\varepsilon /2k_{\rm B}T_{S})} = \frac{\partial f_{i,\rm h}}{\partial [\Delta V_j]},\\
\frac{\partial f_{i,\rm p}}{\partial [\frac{\Delta T_j}{T_{S}}]}&=\delta_{ij} \frac{\varepsilon }{4k_{\rm B}T_{S} \cosh^2(\varepsilon /2k_{\rm B}T_{S})}=-\frac{\partial f_{i,\rm h}}{\partial [\frac{\Delta T_j}{T_{S}}]},
\end{align*}
it is seen that 
in this case the contributing terms to the current are the coherent electron transfer and the crossed Andreev reflection part, i.e. the part $-I^{\alpha}_{j,\rm CET}+I^{\alpha}_{j,\rm CAR}$ of the total current in Eq. (4) of the Letter. 
In the following we
factor out the prefactor $\beta \equiv
\frac{{\rm \delta }^2p}{{\rm \delta }\Omega} \frac{{\cal A}_1^z{\cal A}_2^z}{(2\pi \hbar)^3L^2}$ from Eq. (5) of the Letter.
In a first step we write
\begin{align*}
L^{qT}_{12}&=L^{qT}_{12,\rm CET}+L^{qT}_{12,\rm CAR}\\
&=\beta  \int q\,\frac{-\left[+\varepsilon A_{12,\rm CET}(\varepsilon)\right]+\left[-\varepsilon A_{12,\rm CAR}(\varepsilon)\right]}{4k_{\rm B}T_{S} \cosh^2(\varepsilon /2k_{\rm B}T_{S})}\,d\varepsilon\\
L^{\varepsilon V}_{21}&=L^{\varepsilon V}_{21,\rm CET}+L^{\varepsilon V}_{21,\rm CAR}\\
&= \beta \int \varepsilon\,\frac{-\left[+q A_{21,\rm CET}(\varepsilon)\right]+\left[+q A_{21,\rm CAR}(\varepsilon)\right] }{4k_{\rm B}T_{S} \cosh^2(\varepsilon /2k_{\rm B}T_{S})}\,d\varepsilon ,
\end{align*}
where we introduced $A_{ji,\rm CET/CAR}=(j_{j,\rm CET/CAR}+\tilde j_{j,\rm CET/CAR})/\,\delta f_{i,\rm p/h}$ following the definition of the current.
Let us first compare the terms for the CET contributions.
The Onsager symmetry for the CET contribution follows from
$A_{12,\rm CET}(\varepsilon)=A_{21,\rm CET}(\varepsilon)$. Therefore it suffices to prove this latter equality.
The only terms in the expressions for $A_{12,\rm CET}$ and $A_{21,\rm CET}$ (see Eq. (8) of the Letter) that change under interchanging $1\leftrightarrow2$ are the terms $|v_1u_{12}|$, which change into $|v_2u_{21}|$.
Following the definitions of the Letter, a simple algebra results into
\begin{align*}
v_1u_{12}=\left\{c(1-\Gamma_1\Gamma_2)-i\frac{s}{\omega } [\varepsilon(1+\Gamma_1\Gamma_2)+\Delta(\Gamma_1+\Gamma_2)]\right\}^{-1}.
\end{align*}
This expression is the same as the one resulting from $v_2u_{21}$, as is also obvious from its symmetry with respect to interchanges $1\leftrightarrow2$. Consequently, $|v_1u_{12}|=|v_2u_{21}|$, hence
 $A_{12,\rm CET}=A_{21,\rm CET}$, and therefore 
\begin{align*}
L^{qT}_{12,\rm CET}=L^{\varepsilon V}_{21,\rm CET}. 
\end{align*}

Slightly more involved is the proof for the CAR contributions. 
We will prove that $A_{12,\rm CAR}(\varepsilon)=A_{21,\rm CAR}(-\varepsilon)$.
We will use
the relations $\Gamma_j(\varepsilon;-\delta\varphi_j)=-\Gamma_j^*(-\varepsilon;\delta\varphi_j)$ (the semicolon separates arguments from parameters), which follows from $\gamma_0(\epsilon)=-\gamma_0^\ast(-\epsilon )$ and the definition of the $\Gamma_j(\varepsilon;\delta\varphi_j)$. Note further that from its definition it follows that
$\Gamma_j$ obviously doesn't change under $\uparrow\leftrightarrow\downarrow$. 
We find now from the definitions of $v_1$ and $u_{12}$ and the just mentioned relations
(we use the abbreviations $\delta \varphi_i \equiv \left\{ \delta \varphi_1,\delta \varphi_2\right\}$, $-\delta \varphi_i \equiv \left\{ -\delta \varphi_1,-\delta \varphi_2\right\}$)
\begin{align*}
|v_1u_{12}|^2(\varepsilon;-\delta\varphi_i)&\equiv|v_1(\varepsilon;-\delta\varphi_i)u_{12}(\varepsilon;-\delta\varphi_i)|^2\\&=|v^*_1(-\varepsilon;\delta\varphi_i)u^*_{12}(-\varepsilon;\delta\varphi_i)|^2\\&=|v_1(-\varepsilon;\delta\varphi_i)u_{12}(-\varepsilon;\delta\varphi_i)|^2\\& \equiv|u_1v_{12}|^2(-\varepsilon;\delta\varphi_i)
\end{align*}
From the considerations above for the CET contribution, it follows that 
$ |u_1v_{12}|^2(\varepsilon;\delta\varphi_i) = |u_2v_{21}|^2(\varepsilon;\delta\varphi_i)$.
Furthermore, $|\gamma_0(\varepsilon)|^2=|\gamma_0(-\varepsilon)|^2$.
In order to show the relation
$A_{12,\rm CAR}(\varepsilon)=A_{21,\rm CAR}(-\varepsilon)$ we use Eq. (9) of the Letter.
By introducing $B_{\uparrow\downarrow}= (t_{1\uparrow} t_{2\downarrow})^2 ({r^2_{2\uparrow}} +r^2_{1\downarrow})$ and $B_{\downarrow\uparrow}= (t_{1\downarrow} t_{2\uparrow})^2 ({r^2_{2\downarrow}} +r^2_{1\uparrow})$, we find
\begin{align*}
&A_{12,\rm CAR}(\varepsilon)=\\
&=B_{\uparrow\downarrow}|v_1u_{12}\gamma_0|^2(\varepsilon;\delta\varphi_i)
+B_{\downarrow\uparrow}|v_1u_{12}\gamma_0|^2(\varepsilon;-\delta\varphi_i)\\
&=B_{\uparrow\downarrow}|v_1u_{12}\gamma_0|^2(\varepsilon;\delta\varphi_i)+B_{\downarrow\uparrow}|v_1u_{12}\gamma_0|^2(-\varepsilon;\delta\varphi_i)
\end{align*}
and
\begin{align*}
&A_{21,\rm CAR}(\varepsilon)=\\
&=B_{\downarrow\uparrow}|v_2u_{21}\gamma_0|^2(\varepsilon;\delta\varphi_i)+B_{\uparrow\downarrow}|v_2u_{21}\gamma_0|^2(\varepsilon;-\delta\varphi_i)\\
&=B_{\downarrow\uparrow}|v_1u_{12}\gamma_0|^2(\varepsilon;\delta\varphi_i)+B_{\uparrow\downarrow}|v_1u_{12}\gamma_0|^2(-\varepsilon;\delta\varphi_i)
\end{align*}
and comparing the last lines of these two sets of equations, it follows that
\begin{align*}
A_{21,\rm CAR}(-\varepsilon)&=A_{12,\rm CAR}(\varepsilon ).
\end{align*}
In a last step we find
\begin{align*}
L^{qT}_{12,\rm CAR}&=\beta \int_{-\infty}^{+\infty}\frac{-\varepsilon q A_{12,\rm CAR}(\varepsilon)}{4k_{\rm B}T_{S} \cosh^2(\varepsilon /2k_{\rm B}T_{S})}\,{\rm d}\varepsilon\\&=\beta \int_{-\infty}^{+\infty}\frac{-\varepsilon q A_{21,\rm CAR}(-\varepsilon)}{4k_{\rm B}T_{S} \cosh^2(\varepsilon /2k_{\rm B}T_{S})}\,{\rm d}\varepsilon\\&=\beta \int_{-\infty}^{+\infty}\frac{\varepsilon' q A_{21,\rm CAR}(\varepsilon')}{4k_{\rm B}T_{S} \cosh^2(\varepsilon' /2k_{\rm B}T_{S})} \,{\rm d}\varepsilon' = L^{\varepsilon V}_{21,\rm CAR}\, .
\end{align*}
All remaining Onsager relations are proved analogously [$\gamma_j(\varepsilon;-\delta \varphi_i )=-\gamma_j^\ast(-\varepsilon ; \delta \varphi_i )$ will be needed, and for the Andreev contributions $I_{\rm AR}$ the symmetry requires integration over $\varepsilon$, similar as for $I_{\rm CAR}$ above]. 

Note that for {\it symmetric contact parameters} $B_{\downarrow\uparrow}=B_{\uparrow\downarrow}$, and consequently $A_{12,\rm CAR}(\varepsilon ) $ and $A_{21,\rm CAR}(\varepsilon )$ are both even functions in $\varepsilon $, leading to $L^{qT}_{12,\rm CAR}=L^{\varepsilon V}_{21,\rm CAR}=0$. 
For {\it antiparallel alignment of the magnetization of the two contacts, and otherwise identical contact parameters}, $\Gamma_2(\varepsilon; \delta \varphi_2)=\Gamma_1(\varepsilon; -\delta \varphi_1)=-\Gamma_1^\ast(-\varepsilon; \delta \varphi_1)$, and $|v_1u_{12}|(\varepsilon)=|v_1u_{12}|(-\varepsilon)$, which together with $|\gamma_0(\varepsilon)|=|\gamma_0(-\varepsilon)|$ shows that $L_{12}^{qT}=L_{12}^{\varepsilon V}=L_{21}^{qT}=L_{21}^{\varepsilon V}=0$.

From the above symmetry relations it follows that
for identical contacts and parallel magnetization, there
are only 3 local and 4 nonlocal independent coefficients; for antiparallel magnetization all nonlocal thermoelectric coefficients are zero, and there are 3 local and 2 nonlocal independent coefficients. In the general case, for asymmetric contacts, there are 6 independent local and 4 independent nonlocal coefficients.

\section{Vanishing of pure Andreev reflection contributions to the thermoelectric effects in the clean limit}

Similarly as for the CET and CAR contributions, discussed in the last section,
for the AR contribution one can show that $A_{jj,{\rm AR}}(\varepsilon )=A_{jj,{\rm AR}} (-\varepsilon )$. This follows directly from Eq. (7) of the Letter, and the fact that $\gamma_0(\varepsilon )=-\gamma_0^\ast(-\varepsilon )$ and  $\gamma_j(\varepsilon;\delta -\varphi_j)=-\gamma^\ast_j(-\varepsilon;\delta \varphi_j)|^2$: 
introducing $B'_{j,\uparrow\downarrow}= (t_{j\uparrow} t_{j\downarrow})^2=B'_{j,\downarrow\uparrow}$, we find
\begin{align*}
A_{jj,\rm AR}(\varepsilon)&=
B'_{j,\uparrow\downarrow}\left(|v_1|^2(|\gamma_1|^2+|\gamma_0|^2\right)(\varepsilon;\delta\varphi_i)
\\&+B'_{j,\downarrow\uparrow}\left(|v_1|^2(|\gamma_1|^2+|\gamma_0|^2\right)(\varepsilon;-\delta\varphi_i)\\
&=B'_{j,\uparrow\downarrow}\left\{\left(|v_1|^2(|\gamma_1|^2+|\gamma_0|^2\right)(\varepsilon;\delta\varphi_i) 
\right. \\&
\qquad \quad + \left. \left(|v_1|^2(|\gamma_1|^2+|\gamma_0|^2\right)(-\varepsilon;\delta\varphi_i)\right\},
\end{align*}
leading to $A_{jj,{\rm AR}}(-\varepsilon)=A_{jj,{\rm AR}}(\varepsilon )$. Introducing this into the integrals for 
$L_{jj,{\rm AR}}^{qT}$ and $L_{jj,{\rm AR}}^{\varepsilon V}$ 
shows that both vanish identically:
\begin{align*}
L_{jj,{\rm AR}}^{qT}=L_{jj,{\rm AR}}^{\varepsilon V} =0 .
\end{align*}

\section{Nonlocal origin of all thermoelectric coefficients in the clean limit}

As shown in the previous section, the Andreev reflection contributions to the local thermoelectric effects vanish, such that
local thermoelectric coefficients are solely due to the $I_{\rm R}$ contribution, determined by Eq. (6) of the Letter. The presence of the second spin-polarized contact is, however, crucial. In the absence of a second spin-polarized contact the local thermoelectric coefficients $L_{11}^{qT}$ etc. vanish, as we show in this section. 

This can be understood in the following way. In the absence of a second contact quasiparticles entering through contact 1 can either be Andreev reflected, or scattered from some surface point of the superconductor. After scattering, they will undergo Andreev reflection, with a hole being backscattered
(we assume that any surface point from which direct retro-reflection into the original contact occurs is far away from the first contact on a coherence length scale, and thus contributes negligibly). 
Typical backscattering events are presented in Fig. 1 of this Supplementary Material. 
We show now, that 
backscattering from any specular surface point of the superconductor 
will not lead to any thermopower either. Assume that 
the second contact is specularly reflecting and has zero spin-mixing angle. 
We show that the resulting thermoelectric coefficients vanish. For $\delta \varphi_2=0$ and $t_{2\uparrow}=t_{2\downarrow}=0$ it follows that $\Gamma_2(\varepsilon )=\gamma_0(\varepsilon)$, which leads after some straightforward algebra to $\gamma_1(\varepsilon)=\gamma_0(\varepsilon )$ (meaning that the propagator coming from interface 2 is now equal to one coming from the bulk in the superconductor). Thus,
with 
\begin{align*}
A_{11,\rm R}(\varepsilon)&=
2\left|\frac{r_{1\uparrow }-r_{1\downarrow} e^{i\delta \varphi_1}\gamma_0(\varepsilon)^2}{1-r_{1\uparrow}r_{1\downarrow}e^{i\delta \varphi_1}\gamma_0(\varepsilon)^2}\right|^2
\\&+
2\left|\frac{r_{1\downarrow }-r_{1\uparrow} e^{-i\delta \varphi_1}\gamma_0(\varepsilon)^2}{1-r_{1\uparrow}r_{1\downarrow}e^{-i\delta \varphi_1}\gamma_0(\varepsilon)^2}\right|^2 
\end{align*}
(using $
r_{1\uparrow}-v_1 t_{1\uparrow}^2r_{1\downarrow} e^{i\delta\varphi^{}_1 } \gamma_0\gamma_1 \equiv
v_1[r_{1\uparrow}-r_{1\downarrow} e^{i\delta\varphi^{}_1 } \gamma_0\gamma_1 ]$)
we consider two cases. 
For $|\varepsilon|\ge |\Delta (T)|$ we have a purely real $\gamma_0(\varepsilon)$. In this case,
$\gamma_0(\varepsilon)^2=\gamma_0(-\varepsilon)^2$ and the expression is identical to the one for $A_{11,{\rm R}}(-\varepsilon)$.
On the other hand, for $|\varepsilon|<|\Delta (T)|$ we have $|\gamma_0|=1$, and can thus write $\gamma_0(\varepsilon )=e^{i\Psi(\varepsilon)}$. As the absolute value does not change when multiplying with a phase factor or when taking the complex conjugate, in this case we obtain
\begin{align*}
A_{11,\rm R}(\varepsilon)&=
2\left|\frac{r_{1\uparrow }-r_{1\downarrow} e^{i\delta \varphi_1}\gamma_0(\varepsilon)^2}{1-r_{1\uparrow}r_{1\downarrow}e^{i\delta \varphi_1}\gamma_0(\varepsilon)^2}\right|^2
\\&+
2\left|\frac{r_{1\downarrow }e^{i\delta \varphi_1}\gamma^\ast_0(\varepsilon)^2-r_{1\uparrow} }{1-r_{1\uparrow}r_{1\downarrow}e^{i\delta \varphi_1}\gamma^\ast_0(\varepsilon)^2}\right|^2 
\end{align*}
With $\gamma^\ast_0(\varepsilon)=-\gamma_0(-\varepsilon)$, this expression again is equal to $A_{11,{\rm R}}(-\varepsilon)$.
It follows that for all $\varepsilon$ the relation $A_{11,{\rm R}}(\varepsilon )=A_{11,{\rm R}}(-\varepsilon)$ holds, and introducing this into the integrals for $L_{11,{\rm R}}^{qT}$ and $L_{11,{\rm R}}^{\varepsilon V}$ leads to vanishing results: $L_{11,{\rm R}}^{qT}=L_{11,{\rm R}}^{\varepsilon V}=0$ for $\delta \varphi_2=0$ and $t_{2\uparrow}=t_{2\downarrow }=0$.

Thus, nonzero local thermoelectric coefficients ($L_{11}^{qT}$, $L_{11}^{\varepsilon V}$) require the same nonlocal processes as the nonlocal coefficients ($L_{12}^{qT}$, $L_{12}^{\varepsilon V}$). 

\section{Various definitions of nonlocal thermopower}

In Tab.~\ref{seebeck:nl} we list several possibilities to relate voltage and temperature differences between the two ferromagnets and the superconductor avoiding a control of energy currents. 
We assume a temperature difference $\Delta T_2$ is applied at contact 2 and a voltage $\Delta V_1$ is established at contact 1, which determines the thermopower. The supercurrent flowing out of the superconducting terminal is given by $I_S^q=I_1^q+I_2^q$.
In order to have a clean effect, we require no temperature difference at contact 1. Furthermore, we avoid scenarios where both voltage and current at the same contact should be tuned to zero. We also require that either the current through terminal 1 or through the superconducting terminal should be zero. This gives the combinations shown in Tab.~\ref{seebeck:nl}. 
The experimentally chosen combination depends on the context.

\begin{table}[t]
\begin{ruledtabular}
\begin{tabular}{c||c|c|c|c||c}
        & $I^q_1$       & $I^q_2$       & $\Delta T_1$  & $\Delta \slPhi_2$     & ${\mathcal{S}}\,T_{S}$
        \\\hline\hline
  I     & 0 & free              & 0             &0      &$L^{qT}_{12}/L^{qV}_{11}$\\\hline
  II   &\multicolumn{2}{c|}{$I^q_1+ I^q_2=0$}  & 0             &0      &$\frac{L^{qT}_{12}+ L^{qT}_{22}}{L^{qV}_{11}+ L^{qV}_{21}}$\\\hline
  III     & 0             & 0             & 0             &free   &$\frac{L^{qT}_{12}L^{qV}_{22}-L^{qT}_{22}L^{qV}_{12}}{L^{qV}_{11}L^{qV}_{22}-L^{qV}_{21}L^{qV}_{12}}$
\end{tabular}
\end{ruledtabular}
\caption{\label{seebeck:nl}List of various possibilities to define the nonlocal thermopower $\mathcal{S}\equiv \mathcal{S}_{12}=-\Delta \slPhi_{1}/\Delta T_{2}$ relating the voltage in terminal 1 to the temperature difference in terminal 2 in a three-terminal device. Control of energy currents is avoided.\vspace{-3mm}}
\end{table}

In particular, for combinations I-II at each ferromagnetic terminal either the voltage or the temperature difference is non-zero (but not both), which presents the nonlocal effect in a clean way. For case I a finite supercurrent is flowing out of the superconducting terminal, for cases II and III there is no such supercurrent.
Case III is special in that there are no charge currents flowing in or out of the device. 
In case I and II no local thermopower is present ($\Delta V_2=0$). In case III the nonlocal thermopower ${\cal S}$ is accompagnied by a local thermopower ${\cal S}_2=-\Delta V_2/\Delta T_2$ of value
\begin{figure}[b]
\includegraphics*[width=1.0\linewidth,clip]{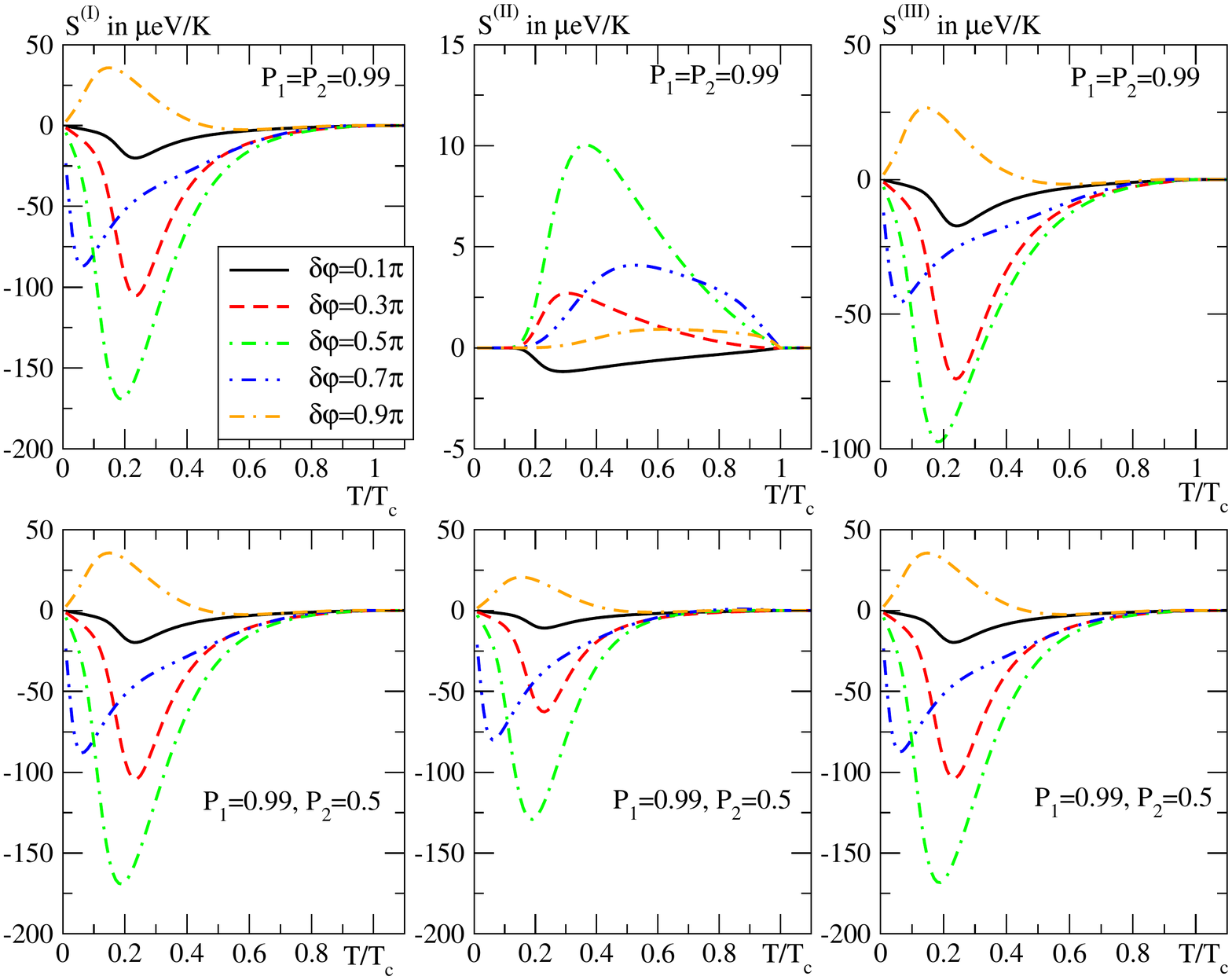}\\
\caption{\label{tp3}
Comparison of the three definitions in Table~\ref{seebeck:nl}. In the left column ${\cal S}^{(I)}$, in the middle column ${\cal S}^{(II)}$, and in the right column ${\cal S}^{(III)}$ are shown. Top row: the spin polarization on both contacts are equal and 99\% (corresponding to a half metal). Bottom row: the spin polarization at contact 2 is only 50 \%: ${\cal P}_1=0.99$ and ${\cal P}_2=0.5$. For all results ${\cal T}_1={\cal T}_2=0.1$ and $\delta \varphi_1=\delta \varphi_2\equiv \delta \varphi$, $\delta \Omega_1=\delta \Omega_2=\pi/20$, $L=0.5\xi_0$.
As can be seen, the symmetric case is special for ${\cal S}^{(II)}$. Otherwise all definitions give qualitatively similar results.
}
\end{figure}
\begin{align*}
{\mathcal{S}_2^{(III)}}\,T_{S} = \frac{L^{qV}_{11}L^{qT}_{22}-L^{qT}_{12}L^{qV}_{21}}{L^{qV}_{11}L^{qV}_{22}-L^{qV}_{21}L^{qV}_{12}}.
\end{align*}
This latter case is interesting from the point of view that a temperature difference at contact 2 causes thermopowers at both contact 1 and contact 2 under condition of no current flow in the device.
These thermopowers fulfill the equation
$
L^{qV}_{12}{\mathcal{S}_2^{(III)}}=
L^{qV}_{11} ({\mathcal{S}_{12}^{(I)}}- {\mathcal{S}_{12}^{(III)}}).
$
In fact, numerical calulation shows (see Fig.~\ref{tp4}) that the sign of ${\mathcal{S}_{12}^{(III)}}$ and ${\mathcal{S}_2^{(III)}}$ is opposite, such that the thermopowers sum up over the two contacts, thus providing effectively a ferromagnet/superconductor/ferromagnet junction with largely enhanced thermopower mediated by spin-polarized Cooper pairs in the superconductor.

In Fig.~\ref{tp3} we compare results for the thermopower using the three cases in Table~\ref{seebeck:nl}. It can be seen that large nonlocal thermopower arises independent on the magnitude of the supercurrent flowing out of the superconducting terminal. 
\begin{figure}[t]
\includegraphics*[width=1.0\linewidth,clip]{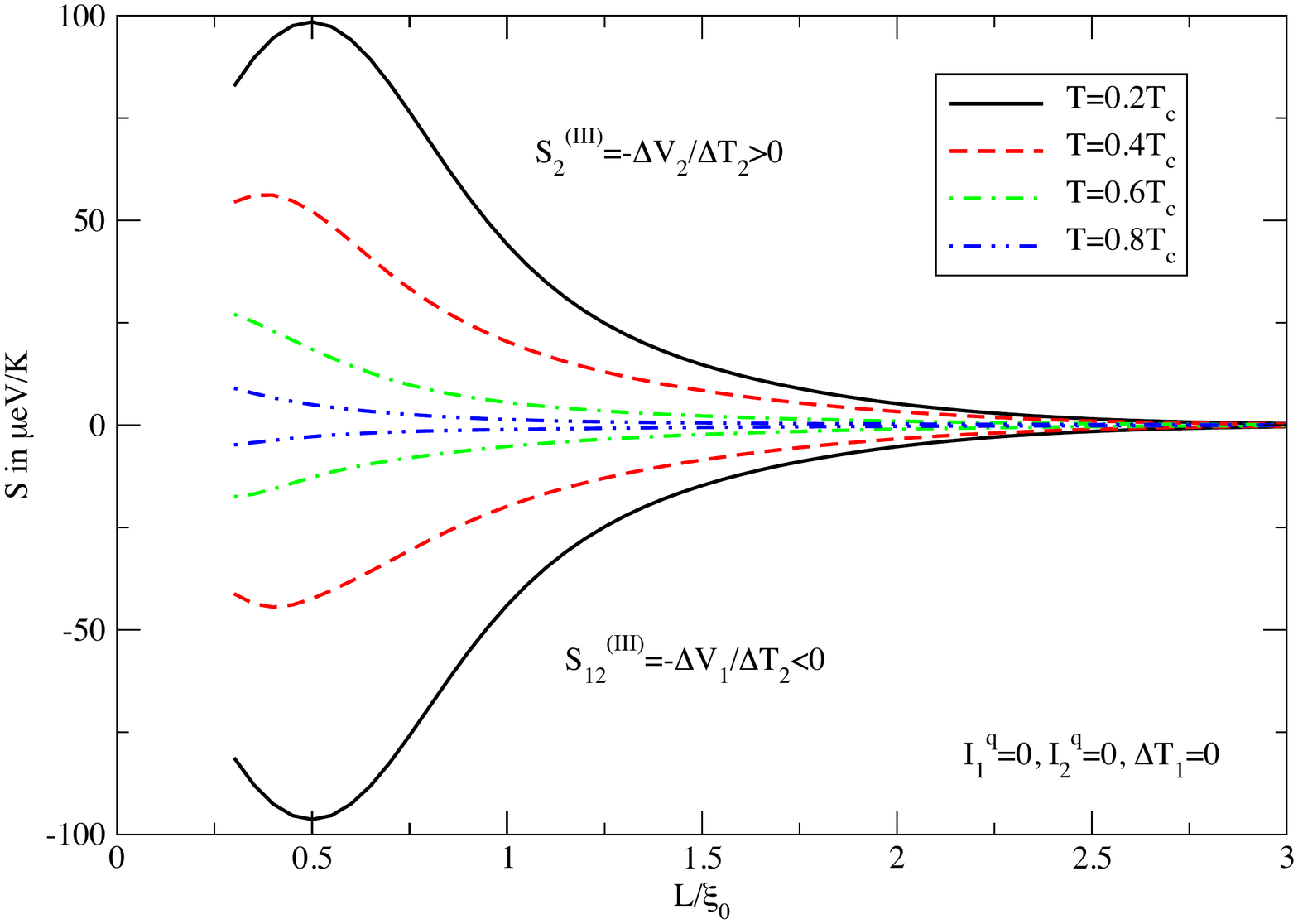}\\
\caption{\label{tp4}
Dependence of thermopower on distance between contacts in the ballistic limit.
The local and nonlocal thermopower are plotted for case III.
Parameters correspond to those in the top row of Fig.~\ref{tp3}, i.e.
${\cal P}_1={\cal P}_2=0.99$, ${\cal T}_1={\cal T}_2=0.1$ and $\delta \varphi_1=\delta \varphi_2= 0.5\pi $, $\delta \Omega_1=\delta \Omega_2=\frac{\pi}{80}(\xi_0/L)^2$.
The voltage in contact 2 is below that in the superconductor, and the one in contact 1 above that in the superconductor. A thermopower between the ferromagnetic leads is established, with no charge currents flowing in the device. The (almost) equal magnitudes are a result of symmetric contact parameters; that is, differing transmissions lead to differing magnitudes.
}
\end{figure}
Fig.~\ref{tp4} shows the dependence of the thermopower on the contact distance.

\bibliographystyle{phaip}